\documentclass[12pt]{article}

\usepackage{amssymb}
\usepackage{amsmath}
\usepackage{siunitx}
\usepackage{geometry}
\usepackage{graphicx}
\usepackage{subcaption}
\usepackage{longtable}
\usepackage{url}
\usepackage{natbib}
\setcitestyle{round,semicolon}

\title{A Multiphase Model of Growth Factor-Regulated Atherosclerotic Cap Formation}
\author{Michael G. Watson, Helen M. Byrne, Charlie Macaskill, Mary R. Myerscough}

\begin{document}

\maketitle
\abovedisplayskip=12pt
\belowdisplayskip=12pt
\setlength{\jot}{2ex}

\begin{abstract}
Atherosclerosis is characterised by the growth of fatty plaques in the inner (intimal) layer of the artery wall. In mature plaques, vascular smooth muscle cells (SMCs) are recruited from the adjacent medial layer to deposit a cap of fibrous collagen over the fatty plaque core. The fibrous cap isolates the thrombogenic content of the plaque from the bloodstream and prevents the formation of blood clots that cause myocardial infarction or stroke. Despite the important protective role of the cap, the mechanisms that regulate cap formation and maintenance are not well understood. It remains unclear why certain caps become stable, while others become vulnerable to rupture.

We develop a multiphase PDE model with non-standard boundary conditions to investigate collagen cap formation by SMCs in response to growth factor signals from the endothelium. Diffusible platelet-derived growth factor (PDGF) stimulates SMC migration, proliferation and collagen degradation, while diffusible transforming growth factor (TGF)-$\beta$ stimulates SMC collagen synthesis and inhibits collagen degradation. The model SMCs respond haptotactically to gradients in the collagen phase and have reduced rates of migration and proliferation in dense collagenous tissue. The model, which is parameterised using a range of \textit{in vivo} and \textit{in vitro} experimental data, reproduces several observations from studies of plaque growth in atherosclerosis-prone mice. Numerical simulations and model analysis demonstrate that a stable cap can be formed by a relatively small SMC population and emphasise the critical role of TGF-$\beta$ in effective cap formation and maintenance. These findings provide unique insight into the cellular and biochemical mechanisms that may lead to plaque destabilisation and rupture. This work represents an important step towards the development of a comprehensive \textit{in silico} plaque.

\end{abstract}

\section{Introduction} \label{sIntro}
Cardiovascular diseases are the leading global cause of mortality \citep{WHO17}. Atherosclerosis, the growth of fat-filled plaques in the walls of arteries, is the primary cause of most cardiovascular disease-related deaths. Unstable atherosclerotic plaques can rupture, and subsequent blood clot formation may occlude blood flow and lead to myocardial infarction or stroke \citep{Hans06}. Smooth muscle cells (SMCs) play an important role in the prevention of these grave clinical outcomes by forming a stabilising cap of fibrous tissue over the lipid-rich core in developing plaques \citep{Alex12}.
To understand how plaques become unstable and dangerous, it is important to understand the behaviour of plaque SMCs and the corresponding mechanisms of fibrous cap synthesis or degradation. Despite extensive experimental investigation, the \emph{in vivo} dynamics of cap formation remain poorly understood.

The absence of a dynamic understanding of cap formation and maintenance can be largely attributed to the slow rate of plaque development. In humans, atherosclerosis can begin in childhood and plaques may take several decades to progress towards a dangerous, unstable state \citep{Lusi00}. Even in the apolipoprotein-E (ApoE) deficient mouse (the most common animal model for experimental atherosclerosis studies), plaques require a period of several months to grow to an advanced stage. Consequently, observations of plaque tissue resected from experimental mice are typically made at distinct (or even single) time points. Mathematical modelling provides a powerful tool to study the dynamic mechanisms that underlie these patchy experimental observations. In this paper, we use the multiphase framework developed in \citet{Wats18} and build a comprehensive new model to study the formation of the protective fibrous cap by atherosclerotic plaque SMCs.

\begin{figure}
		\centering		
  		\includegraphics[height=6cm]{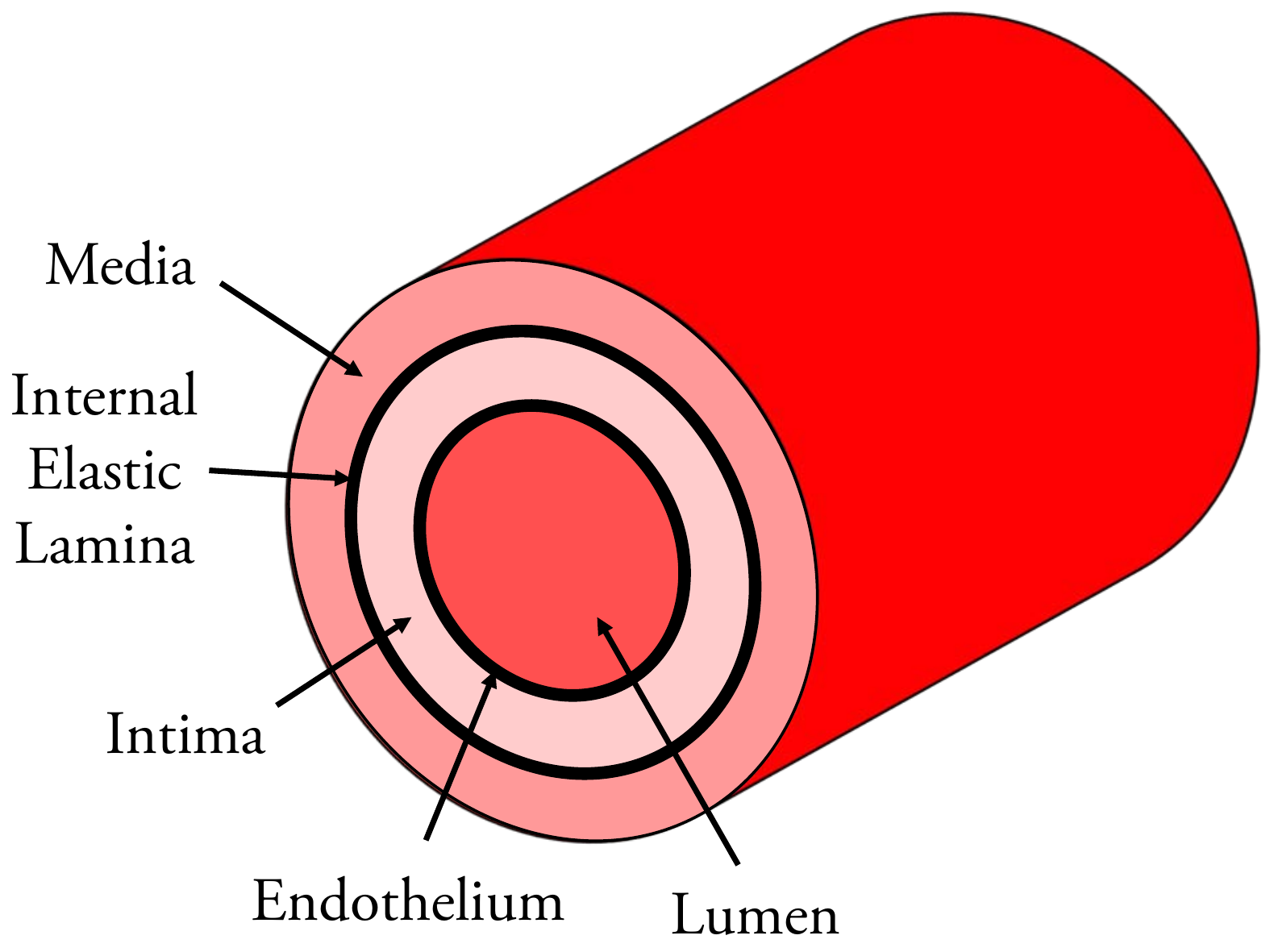}
		\caption{Schematic diagram of a cross-section through the inner artery wall (layer widths not to scale and outer adventitial layer not shown). Atherosclerotic plaques develop in the \emph{intima}, which is separated from the blood flow in the \emph{lumen} by a thin layer of cells called the \emph{endothelium}. Deeper in the wall, the intima is separated from the \emph{media} by a membrane called the \emph{internal elastic lamina}.}\label{ArtSchem}
	\end{figure}

Atherosclerotic plaques develop in the narrow intimal layer of the artery wall (Figure~\ref{ArtSchem}). The intima is located between the endothelium (a thin sheet of endothelial cells that lines the vessel lumen) and the media, which houses several striated layers of quiescent (contractile) SMCs. The intima and the media are separated by a dense tissue membrane known as the internal elastic lamina (IEL). The outermost layer of the artery wall beyond the media is called the adventitia. Plaque formation is initiated by blood-borne low-density lipoproteins (LDL), which accumulate in the intima and become oxidised or modified in different ways. The presence of modified LDL triggers the recruitment of immune cells (monocytes) from the bloodstream by a process of transendothelial migration. These monocytes differentiate into macrophages, which can consume the lipid and remove it from the intima by migration to the adventitial lymphatics \citep{Moor13}. However, when lipid-filled macrophages (known as foam cells) die within the plaque, they release their lipid content and other cellular debris. This reinforces the immune response and can lead to a cycle of chronic inflammation that promotes plaque growth and, ultimately, the development of necrotic tissue.  

\begin{figure}
		\centering		
  		\includegraphics[height=6cm]{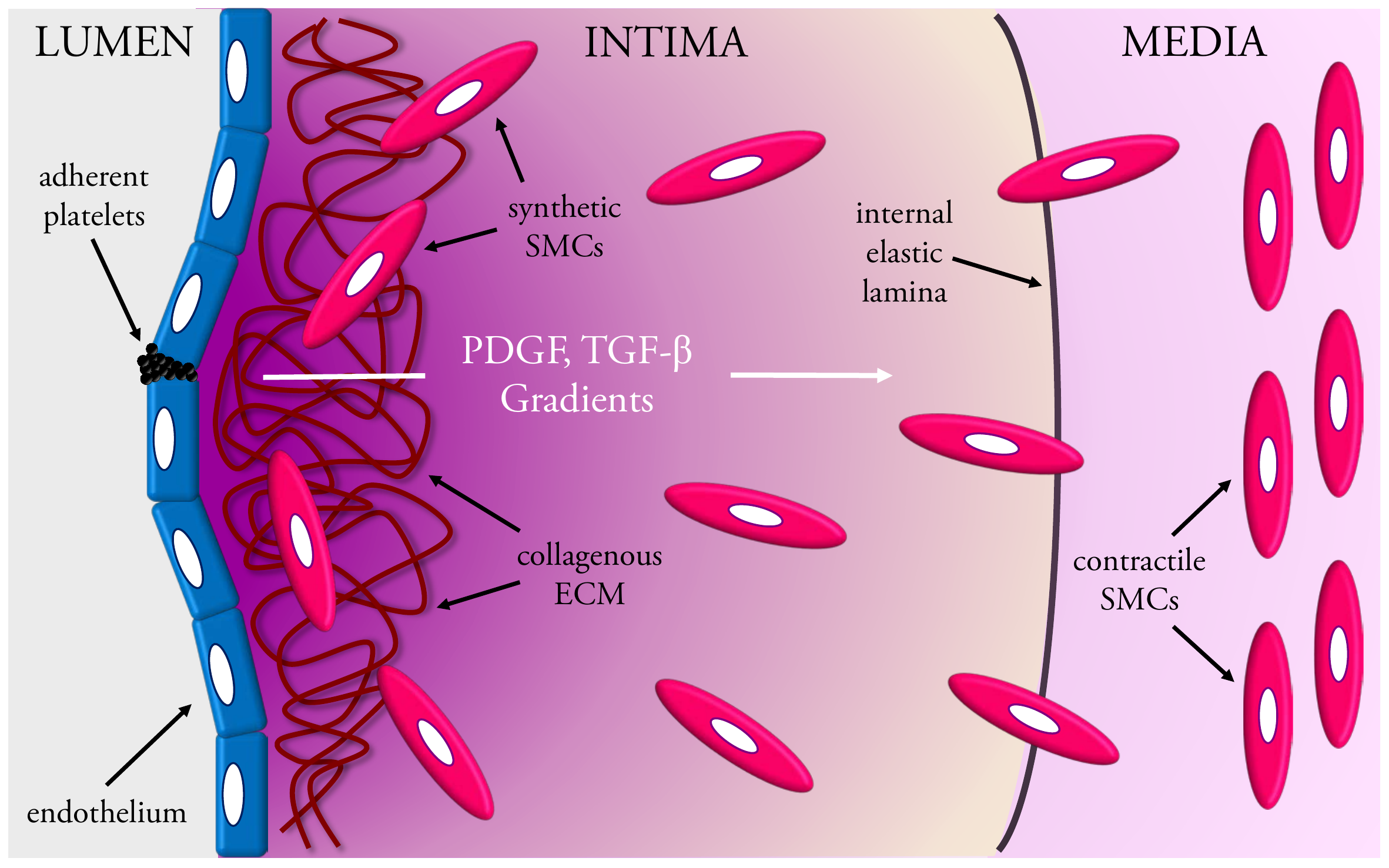}
		\caption{Schematic diagram of the key processes in atherosclerotic cap formation. Lipoprotein accumulation in the intima triggers an immune response that leads to intimal growth and endothelial disruption. Diffusible growth factors PDGF and TGF-$\beta$ are released from sites of injury by endothelial cells and adherent platelets. Contractile SMCs in the media are stimulated to adopt a synthetic state and respond chemotactically to PDGF by migrating through the internal elastic lamina. Once inside the plaque, synthetic SMCs remodel the existing ECM and are stimulated by TGF-$\beta$ to deposit a dense cap of collagenous tissue adjacent to the endothelium.}\label{CapSchem}
	\end{figure}

At an intermediate stage of plaque formation, medial SMCs adopt an active (synthetic) state and migrate through the IEL into the intima. SMC activation is believed to be initiated by an injured endothelium, which may suffer mechanical disruption as the intimal layer expands to accommodate the infiltration of lipid and immune cells \citep{Fagg84}. Once inside the plaque, SMCs replace the existing intimal extracellular matrix (ECM) with a dense matrix of fibrillar collagens \citep{Adig09}. SMCs are subject to an abundance of signalling cues in the plaque, but two chemicals known to be particularly critical to cap formation are platelet-derived growth factor (PDGF) and transforming growth factor-$\beta$ (TGF-$\beta$). PDGF plays an important role as a SMC mitogen and chemoattractant \citep{Ruth97, Sano01}, while TGF-$\beta$ enhances plaque stability by stimulating SMC collagen production \citep{Mall01, Lutg02}. PDGF and TGF-$\beta$ can be produced by a variety of plaque cells, but circulating platelets that adhere to sites of endothelial injury are believed to be a significant source of both growth factors \citep{Ross99, Toma12}. TGF-$\beta$ is secreted as a latent complex and the active component of TGF-$\beta$ must be liberated before it can bind to target cells \citep{Sing06}. Liberation of active TGF-$\beta$ typically requires chemical or mechanical stimulation, and TGF-$\beta$ derived from platelets is believed to be activated by shear forces in the bloodstream \citep{Aham08}. A schematic diagram of the cap formation process is provided in Figure~\ref{CapSchem}.

Recent advances in cell lineage tracing techniques have begun to reveal new insights into the behaviour of plaque SMCs. \citet{Chap16} and \citet{Jaco17} have performed studies using so-called \emph{Confetti} transgenic mice, where individual medial SMCs can be labelled by inducing the unique expression of one of four possible fluorescent proteins in each cell. Both of these studies have identified that SMC populations in advanced plaques consist of several large monochromatic patches, indicating that plaque SMCs are oligoclonal and derived from only a limited number of medial progenitor cells. These results are significant because they highlight, in the ApoE mouse at least, that proliferation makes a much larger contribution to SMC accumulation in plaques than previously thought.

These studies provide an exciting new window into the cellular mechanisms of fibrous cap formation, and have significant potential to advance current understanding of atherosclerosis progression. However, much is still unknown about the factors that underlie the formation and maintenance of the collagenous cap and, in particular, the reasons why certain caps remain thick and stable whilst others become thin and vulnerable to rupture. Several possible modes of long-term cap degradation have been proposed in the literature. These include an increase in SMC death --- possibly due to cellular senescence \citep{Wang15} --- or an increase in collagen destruction due to either elevated matrix metalloproteinase (MMP) production by inflammatory cells \citep{Hans15} or reduced SMC sensitivity to TGF-$\beta$ signalling \citep{Chen07, Veng15}. The model that we develop in this paper addresses these gaps in biological understanding by establishing a framework to simulate cap formation dynamics and assess the deleterious impact of a variety of ECM degradation mechanisms.

We have recently published a foundational two-phase PDE model that studied the migration and proliferation of media-derived SMCs in response to an endothelium-derived source of diffusible PDGF \citep{Wats18}. The model domain was taken to be a one-dimensional cross-section through the diseased intima and the flux of SMCs and PDGF into the plaque were captured by a set of non-standard boundary conditions. The model did not explicitly consider the synthesis of a fibrous cap, but the approach did provide several interesting insights into the mechanisms that regulate SMC migration to the cap region. In particular, the model predicted that SMC recruitment from the media is likely to be a rate-limiting factor for SMC accumulation in plaques --- an observation that is supported by the recent lineage tracing studies of \citet{Chap16} and \citet{Jaco17}. In the current paper, we build upon this earlier model by incorporating a profile of diffusible, endothelium-derived TGF-$\beta$ and an explicit representation of collagenous ECM remodelling. This allows us to perform a detailed investigation of growth factor-stimulated fibrous cap formation by plaque SMCs. Note that the new model is more than just an elementary extension of the earlier approach. We assume that the SMCs and the nascent ECM are coupled by a nonlinear mechanical feedback that allows us to investigate the influence of factors including haptotactic SMC migration, which may play an important role in the cap formation process \citep{Nels96, Lope13}.

Interest in mathematical modelling of the cell activity in atherosclerosis has grown over recent years. The majority of studies published to date have focussed on modelling the inflammatory response that characterises the early stages of plaque development \citep{ElKh07, Papp08, Cohe14, Chal15}, where several groups have coupled their models to detailed representations of blood flow and/or intimal growth \citep{Bule12, Fili13, Isla16, Yang16, Bhui17}. In contrast, only a handful of models have been proposed to study events in the later stages of plaque progression. \citet{McKa04} performed foundational work in this area by proposing a PDE model that included both plaque SMC recruitment and subsequent collagen deposition. The only other model to consider collagen synthesis by invading SMCs is that of \citet{Cill14}, who developed a comprehensive 3D model of blood flow, transmural transport and plaque growth. Neither of these modelling approaches, however, provide an adequate description of the cap formation process. Several other models of plaque progression \citep{Post07, Frie15} --- including studies of PDGF-induced intimal thickening \citep{Fok12} and intraplaque haemorrhage \citep{Guo18} --- have considered SMCs independently of their collagen-synthesising activity. Interested readers are referred to \citet{Part16} for a comprehensive review of mathematical and computational approaches to atherosclerosis modelling.

Beyond atherosclerosis, a variety of mathematical models have been developed to study the response of vascular SMCs to other sources of endothelial injury. One area that has attracted significant research interest is in-stent restenosis --- the rapid recurrence of a narrowed lumen after surgical deployment of an artery-widening stent. Mechanical stresses imposed by the stent can locally denude the endothelium and elicit an intense healing response that involves rapid proliferation of medial SMCs and significant neointima formation. Discrete and continuous models of SMC behaviour during in-stent restenosis have been developed by several research groups \citep{Lall06, Evan08, Zahe14, Tahi15}. Researchers have also developed models of the vascular SMC response to surgical interventions such as vein grafting \citep{Budu08, Garb17} and blood filter insertion \citep{Nico15}. The tissue repair carried out by artery wall SMCs in response to vascular injury also shares several similarities with the process of dermal wound healing. In dermal wounds, fibroblasts take the role of the SMC and migrate to the wound site to regenerate the damaged collagenous tissue. Wound fibroblasts are known to be stimulated by growth factors including PDGF and TGF-$\beta$, and the corresponding implications for healing have been studied in a variety of modelling frameworks \citep{Olse95, Cobb00, Haug06, McDo06, Cumm10, Meno12}.

The model of cap formation that we present in this paper is inspired by the multiphase theory developed in \citet{Byrn04}, \citet{Lemo06} and \citet{Asta08}. Multiphase models have been widely developed to study articular cartilage (reviewed in \citet{Klik16}), tumour growth \citep{Prez09, Hubb13} and tissue engineering applications \citep{ODea13, Pear14}, but our previous study \citep{Wats18} was the first application of the approach in atherosclerosis. Other existing studies of plaque development have predominantly utilised reaction-diffusion equations to model the spatio-temporal evolution of cells and tissues in the plaque \citep{McKa04, ElKh07, Fili13, Cill14, Chal15, Chal17}. However, multiphase models can provide a more detailed representation of plaque formation dynamics because they provide a natural framework to account for volume exclusion and mechanical interactions between individual plaque constituents. We continue to develop our multiphase approach in this study because we firmly believe that both volume exclusion and mechanical effects can play a significant role in the development of plaque spatial structure.

In the next section, we derive the model equations and introduce the model parameterisation. Where possible, we base our assumptions and our parameter selections on observations of plaque growth in the ApoE mouse, which is the most well-characterised animal model of \emph{in vivo} atherosclerosis in the experimental literature \citep{Getz12}. We present the model results in Section~\ref{sResults}, where we report some interesting analysis before performing a series of numerical simulations. We use these results to investigate the dynamic cellular, biochemical and mechanical mechanisms that regulate cap formation and aim to identify the factors that are most important to the development and maintenance of plaque stability. Finally, we conclude with a discussion of the outcomes of the study in the context of both computational and experimental atherosclerosis research.

\section{Model Formulation and Parameterisation} \label{sMFP}
The plaque tissue in the model is represented as a mixture of three distinct phases: a cellular phase that comprises matrix-synthesising SMCs, an ECM phase that comprises collagen-rich fibrous tissue, and a generic phase that is assumed to comprise the remaining plaque constituents (foam cells, extracellular lipids, interstitial fluid, etc.). We derive mass and momentum balance equations for each phase, and the equations are closed by imposing suitable constitutive assumptions and conditions for the transfer of mass and momentum between each pair of phases. Equations for the concentrations of diffusible PDGF and (active) TGF-$\beta$ are also included in the model. We assume that PDGF promotes a migroproliferative SMC phenotype, while TGF-$\beta$ promotes a matrigenic SMC phenotype \citep{Alex12}. These distinct SMC phenotypes are not explicitly represented in the model. Instead, we assume that the SMC phase contains a continuum of different phenotypes whose average behaviour is regulated by the local growth factor concentrations. We assume that both PDGF and TGF-$\beta$ reside exclusively in the generic tissue phase, and that they exert influence on the SMC and ECM phases via stress and/or mass transfer terms that are functions of their respective concentrations.

We model the above system on a one-dimensional Cartesian domain that represents a cross-section of diseased arterial intima far from the edges of the plaque. We assume that the domain is bounded by the endothelium at $x=0$ and by the IEL at $x=L$. The dependent variables are therefore functions of time $t\geqslant0$ and position $x\in[0,L]$. We denote the volume fractions of the SMC, ECM and generic tissue phases by $m\left(x,t\right)$, $\rho\left(x,t\right)$ and $w\left(x,t\right)$, and their corresponding intraphase stresses by $\tau_m\left(x,t\right)$, $\tau_\rho\left(x,t\right)$ and $\tau_w\left(x,t\right)$, respectively. We assume that the ECM phase takes the form of a rigid scaffold and the ECM phase therefore has zero velocity for all $x$ and $t$. We denote the (non-zero) velocities of the SMC and generic tissue phases by $v_m\left(x,t\right)$ and $v_w\left(x,t\right)$. The interstitial fluid pressure is denoted by $p\left(x,t\right)$, and the concentrations of PDGF and TGF-$\beta$ in the generic tissue phase are denoted by $P\left(x,t\right)$ and $T\left(x,t\right)$, respectively.

\subsection{SMC, ECM and Generic Tissue Phases} \label{ssSEO}
In this section we use the principles of mass and momentum conservation to derive equations that describe the SMC, ECM and generic tissue phase dynamics in response to diffusible PDGF and TGF-$\beta$ in the plaque.

\subsubsection{Mass Balance Equations}
Assuming that all three phases share the same constant density, the mass balance equations for $m$, $\rho$ and $w$ can be expressed as follows:
	\begin{align}
		\frac{\partial m}{\partial t} + \frac{\partial}{\partial x}\left(v_m m\right) &= S_m, \label{mmas}\\
		\frac{\partial \rho}{\partial t} &= S_\rho, \label{rmas}\\
		\frac{\partial w}{\partial t} + \frac{\partial}{\partial x}\left(v_w w\right) &= -\left(S_m+S_\rho\right). \label{wmas}
	\end{align}
Here, the functions $S_m$ and $S_\rho$ represent the net rates of SMC and ECM production. We have assumed in equation~(\ref{wmas}) that there is no local source or sink of material and that mass simply transfers from one phase to another as SMCs proliferate or die, and as ECM is synthesised or degraded. We further assume that there are no voids in the plaque tissue, and the three volume-occupying phases therefore satisfy the condition:
	\begin{equation} \label{nvoi}
		m + \rho + w = 1.
	\end{equation}

\subsubsection{Momentum Balance Equations and Constitutive Assumptions}
Neglecting inertial effects and assuming that no external forces act on the system, the momentum balance equations for $m$, $\rho$ and $w$ reduce to a balance between the intraphase and interphase forces that act on each phase:
	\begin{align}
		\frac{\partial}{\partial x}\left(\tau_m m\right) &= F_{\rho m}+F_{w m}, \label{mmom} \\
		\frac{\partial}{\partial x}\left(\tau_\rho \rho\right) &= F_{m \rho}+F_{w \rho}, \label{rmom} \\
		\frac{\partial}{\partial x}\left(\tau_w w\right) &= F_{m w}+F_{\rho w}, \label{wmom}
	\end{align}
where $F_{ij} \left(= -F_{ji}\right)$ denotes the force exerted by phase $j$ on phase $i$.

Following \citet{Lemo06}, we assume that the $F_{ij}$ terms in equations~(\ref{mmom})--(\ref{wmom}) are comprised of interphase pressure and interphase drag components:
	\begin{align}
		F_{m \rho} &= \left(p + \psi\right) \rho \, \frac{\partial m}{\partial x} - \left(p + \psi\right) m \, \frac{\partial \rho}{\partial x} - k_{m \rho} m \rho v_m, \label{Fmr} \\
		F_{m w} &= p w \, \frac{\partial m}{\partial x} - p m \, \frac{\partial w}{\partial x} + k_{m w} m w \left(v_w - v_m\right), \label{Fmw} \\
		F_{\rho w} &= p w \, \frac{\partial \rho}{\partial x} - p \rho \, \frac{\partial w}{\partial x} + k_{\rho w} \rho w v_w. \label{Frw}
	\end{align}
The first term on the right hand side of each of equations~(\ref{Fmr})--(\ref{Frw}) describes the interfacial force exerted by phase $j$ on phase $i$. This force is assumed to be proportional to the interphase pressure between the phases and to the degree of contact of phase $j$ with phase $i$. The force is assumed to act in the direction of increasing interfacial contact and, on the macroscale, this contributes a further term proportional to the gradient in phase $i$. The second term on the right hand side of each equation contributes a corresponding reaction force, which is exerted by phase $i$ on phase $j$. Note that we have included an additional (contact-dependent) interphase pressure contribution $\psi \equiv \psi\left(m,\rho\right)$ in the equation for $F_{m \rho}$. This additional pressure is included to account for traction forces that are generated as the SMCs translocate through the ECM. We assume that traction forces between the other pairs of phases in the system are negligible in comparison, so the equations for $F_{m w}$ and $F_{\rho w}$ include only the contact-independent pressure $p$ in their interphase pressure terms. The final term on the right hand side of each equation represents the interphase drag, which we assume to be proportional to the relative velocities of phases $i$ and $j$ with constant of proportionality $k_{ij} \left(= k_{ji}\right)$. We further assume that the drag depends linearly on the volume fraction of each phase $i$ and $j$ so that no drag can occur in the absence of either phase. 

The generic tissue phase contains several active constituents such as foam cells and macrophages, but we shall assume here that it behaves as an inert isotropic fluid. We make this assumption in order to focus on how the SMC phase interacts with other key factors in the plaque environment. We shall assume that the SMC phase displays more complex behaviour than the generic tissue phase and, in particular, that SMCs alter their motility in response to both the local PDGF concentration and the local ECM volume fraction. Neglecting viscous effects, and combining the approaches of \citet{Byrn04} and \citet{Lemo06}, we adopt the following models for the intraphase stress terms in equations~(\ref{mmom})--(\ref{wmom}):
	\begin{align}
		\tau_m &= - \left[\,p + \rho \psi\left(m,\rho\right) + \Lambda \! \left(P\right)\,\right], \label{taum} \\
		\tau_\rho &= - \left[\,p + m \psi\left(m,\rho\right)\,\right], \label{taur} \\
		\tau_w &= - \, p. \label{tauw}
	\end{align}
In equation~(\ref{taum}), the extra pressures $\Lambda \! \left(P\right)$ and $\rho \psi\left(m,\rho\right)$ describe the respective influences of the PDGF concentration and the ECM volume fraction on the pressure in the SMC phase. The ECM-derived extra pressure term is proportional to the interphase pressure contribution $\psi$ and is weighted according to the degree of contact between the SMC and ECM phases. We assume that the SMC phase has a complementary influence on the pressure in the ECM phase, and this is captured by the extra pressure term $m \psi\left(m,\rho\right)$ in equation~(\ref{taur}). We define the functions $\Lambda$ and $\psi$ to be \citep{Byrn04, Lemo06}:
	\begin{align}
		\Lambda \! \left(P\right) &= \frac{\chi_P}{1 + {\left(\kappa P \right)}^{n_P}}, \label{lamb} \\
		\psi \left(m,\rho\right) &= - \chi_{\rho} + \frac{\delta m^{n_{\rho}}}{\left(1 - m - \rho\right)^{n_{\rho}}}, \label{psi}
	\end{align}
where $\chi_P$, $\kappa$, $n_P$, $\chi_{\rho}$, $\delta$ and $n_{\rho}$ are positive parameters. Equation~(\ref{lamb}) is deliberately chosen to be a decreasing function of $P$ as this ensures that chemotactic SMC migration up PDGF gradients will provide stress relief in the SMC phase. The first term in equation~(\ref{psi}), which ensures that $\psi$ will be negative for sufficiently small $m$, reflects an assumption that the SMCs have an affinity for the collagenous ECM. Note that this adhesion mechanism introduces the potential for haptotactic SMC migration in the presence of ECM gradients. The second term in equation~(\ref{psi}) contributes a net repulsion on the SMC phase that increases significantly as $m$ or $\rho$ become sufficiently large.

\subsubsection{Mass Transfer Terms}
We assume that the SMC phase source term $S_m$ in equation (\ref{mmas}) includes both a linear death term and a proliferation term that requires the uptake of material from the generic tissue phase. We assume that the net SMC proliferation rate is proportional to the sum of a constant baseline mitosis rate $r_m$ and an additional rate that depends on the local PDGF concentration. PDGF is a well-known mitogen for SMCs, and \emph{in vitro} evidence suggests that SMC proliferation can be significantly upregulated with saturating kinetics in the presence of increasing PDGF concentrations \citep{Munr94}. Combining these considerations, we define:
	\begin{equation} \label{sm}
		S_m = r_m m w \left[ \, 1 + \frac{A_m P}{c_m + P} \, \right] - \beta_m m,
	\end{equation}
where $A_m$ quantifies the maximum possible PDGF-stimulated increase in the baseline SMC mitosis rate, $c_m$ represents the PDGF concentration at which the half-maximal increase in the baseline SMC mitosis rate occurs, and $\beta_m$ denotes the rate of SMC death. We remark here that, while normal vascular SMCs are known to be growth inhibited by TGF-$\beta$, \emph{in vitro} experiments have shown that atherosclerotic plaque-derived SMCs retain their proliferative capacity in the presence of TGF-$\beta$ \citep{McCa95}.

We define the ECM phase source term $S_{\rho}$ in equation~(\ref{rmas}) by assuming that the collagenous matrix is continuously remodelled by the cells that occupy the plaque. Based on a variety of experimental observations, we assume that both PDGF and TGF-$\beta$ are important mediators of this remodelling process. We propose the following form for the ECM phase source term, where we assume that ECM is synthesised by SMCs but is also degraded by both SMCs and plaque immune cells: 
	\begin{equation} \label{sr}
		S_{\rho} = r_s m w \left[ \, 1 + \frac{A_s T}{c_s + T} \, \right] - r_d m \rho \left[ \, 1 + \frac{A_d P}{\left(c_d + P\right) \left(1 + \gamma_d T\right)} \, \right] - \beta_\rho \rho w \left[ \, \frac{1+\varepsilon \, \gamma_{\rho} T}{1+\gamma_{\rho} T} \, \right].
	\end{equation}

The first term in equation~(\ref{sr}) represents the net rate of ECM synthesis by the SMC phase, which we assume to be proportional to the sum of a constant baseline ECM production rate $r_s$ and an additional rate that depends on the local TGF-$\beta$ concentration. The role of TGF-$\beta$ as a stimulant for SMC collagen production in plaques has been clearly demonstrated by \emph{in vivo} studies \citep{Mall01, Lutg02} and, following \emph{in vitro} observations \citep{Kubo03}, we assume that the rate of ECM synthesis increases with saturating kinetics in response to increasing $T$. The parameter $A_s$ quantifies the maximum possible TGF-$\beta$-stimulated increase in the baseline ECM synthesis rate, while $c_s$ denotes the TGF-$\beta$ concentration at which the half-maximal increase in the baseline ECM synthesis rate occurs. Note that the ECM synthesis rate is also proportional to $w$ because we assume that ECM synthesis by SMCs again requires the uptake of material from the generic tissue phase.

The second and third terms in equation~(\ref{sr}) represent the respective net rates of ECM degradation by plaque SMCs and by plaque immune cells, both of which are known to produce MMPs. We assume that the net rate of ECM degradation by SMCs is proportional to the sum of a constant baseline degradation rate $r_d$ and an additional rate that depends on the local concentration of both growth factors. \emph{In vitro} studies indicate that SMCs upregulate their production of MMP-2 and MMP-9 in response to stimulation with PDGF, but also that co-stimulation with TGF-$\beta$ can inhibit this response \citep{Borr06, Risi10}. We represent this behaviour with a simple functional form where $A_d$ quantifies the maximum possible PDGF-stimulated increase in the baseline ECM degradation rate, $c_d$ denotes the PDGF concentration at which the half-maximal PDGF-stimulated increase in the baseline ECM degradation rate occurs and $\gamma_d$ quantifies the extent of inhibition of PDGF-stimulated ECM degradation by TGF-$\beta$. Experimental observations also indicate that TGF-$\beta$ can protect the plaque ECM from degradation by MMP-9-producing macrophages \citep{Vada01, Ogaw04}. Assuming that macrophages and other inflammatory cell types comprise a certain fraction of the generic tissue phase, we capture this behaviour in the final term by assuming that the net rate of immune cell ECM degradation is proportional to both $w$ and a decreasing function of $T$. The parameter $\beta_{\rho}$ denotes the maximum possible rate of immune cell ECM degradation, which we assume to be related to the extent of inflammation in the plaque tissue. The dimensionless parameter $\varepsilon \leqslant 1$ quantifies the minimum possible rate of TGF-$\beta$-inhibited immune cell ECM degration (i.e.\ $\varepsilon \beta_{\rho}$), while $\gamma_{\rho}$ quantifies the rate at which immune cell ECM degradation decreases with increasing $T$.

\subsection{Diffusible Growth Factors} \label{ssGF}
In this section we derive reaction-diffusion equations that describe how the concentrations of diffusible PDGF and TGF-$\beta$ evolve in the plaque tissue. In addition to the equations presented below, the model also includes a source term for each growth factor at the endothelium ($x=0$) and sink term for each growth factor at the media ($x=L$). These additional terms are given by boundary conditions and will be discussed in Section~\ref{ssBIC}.

\subsubsection{Mass Balance Equations}
We assume that the characteristic timescales of PDGF and TGF-$\beta$ diffusion are significantly shorter than the characteristic timescales of SMC migration and ECM remodelling. We therefore neglect advective growth factor transport in the generic tissue phase and assume the following quasi-steady state mass balance equations for $P$ and $T$:
	\begin{align}
		0 &= D_P \, \frac{\partial}{\partial x}\left[ \, w \, \frac{\partial P}{\partial x} \, \right] + S_P, \label{cpmas} \\
		0 &= D_T \, \frac{\partial}{\partial x}\left[ \, w \, \frac{\partial T}{\partial x} \, \right] + S_T. \label{ctmas}
	\end{align}
Here, $D_P$ and $D_T$ are constant chemical diffusion coefficients, while $S_P$ and $S_T$ denote local sink terms for PDGF and TGF-$\beta$, respectively. Note that, following \citet{Asta08}, we have assumed that the diffusive flux of both chemicals is modulated by the factor $w$. This implies that the capacity for growth factor transport through the intimal tissue will reduce as SMCs and ECM accumulate in the plaque.

\subsubsection{Source Terms}
We assume that both growth factors are taken up by SMCs in the plaque, and also that both growth factors undergo natural decay. These assumptions lead to the sink terms:
	\begin{align}
		S_P &= - \eta_P m w P - \beta_P w P, \label{scp} \\
		S_T &= - \eta_T m w T - \beta_T w T, \label{sct}
	\end{align}
where $\eta_P$, $\eta_T$, $\beta_P$ and $\beta_T$ denote the rates of PDGF uptake by SMCs, TGF-$\beta$ uptake by SMCs, PDGF decay and TGF-$\beta$ decay, respectively. Note that we include a factor of $w$ in each term to ensure that neither uptake nor decay of either growth factor can occur in the absence of the generic tissue phase.

\subsection{Model Simplification} \label{ssMS}
In this section, we show how the three-phase model developed in Section \ref{ssSEO} can be reduced to a system of two coupled nonlinear equations for the SMC and ECM phases. In all that follows, we shall assume that the volume fractions of all three phases $m$, $\rho$ and $w$ are strictly positive for all $x$ and $t$.

Using equation~(\ref{nvoi}) to replace all occurrences of $w$ with the equivalent expression $1 - m - \rho$, the sum of the mass balance equations~(\ref{mmas})--(\ref{wmas}) reduces to:
	\begin{equation} \label{summa}
		\frac{\partial}{\partial x} \Big[ v_m m + v_w \left(1 - m - \rho \right) \Big] = 0.
	\end{equation}
Assuming a zero-flux condition for the SMC phase at the endothelium (i.e.\ $v_m = v_w = 0$ at $x = 0$), integration of equation~(\ref{summa}) with respect to $x$ gives the following expression for the velocity of the generic tissue phase:
	\begin{equation} \label{wvel}
		v_w = -v_m \left( \frac{m}{1 - m - \rho} \right).
	\end{equation}

\begin{figure}
		\centering		
  		\includegraphics[height=8cm]{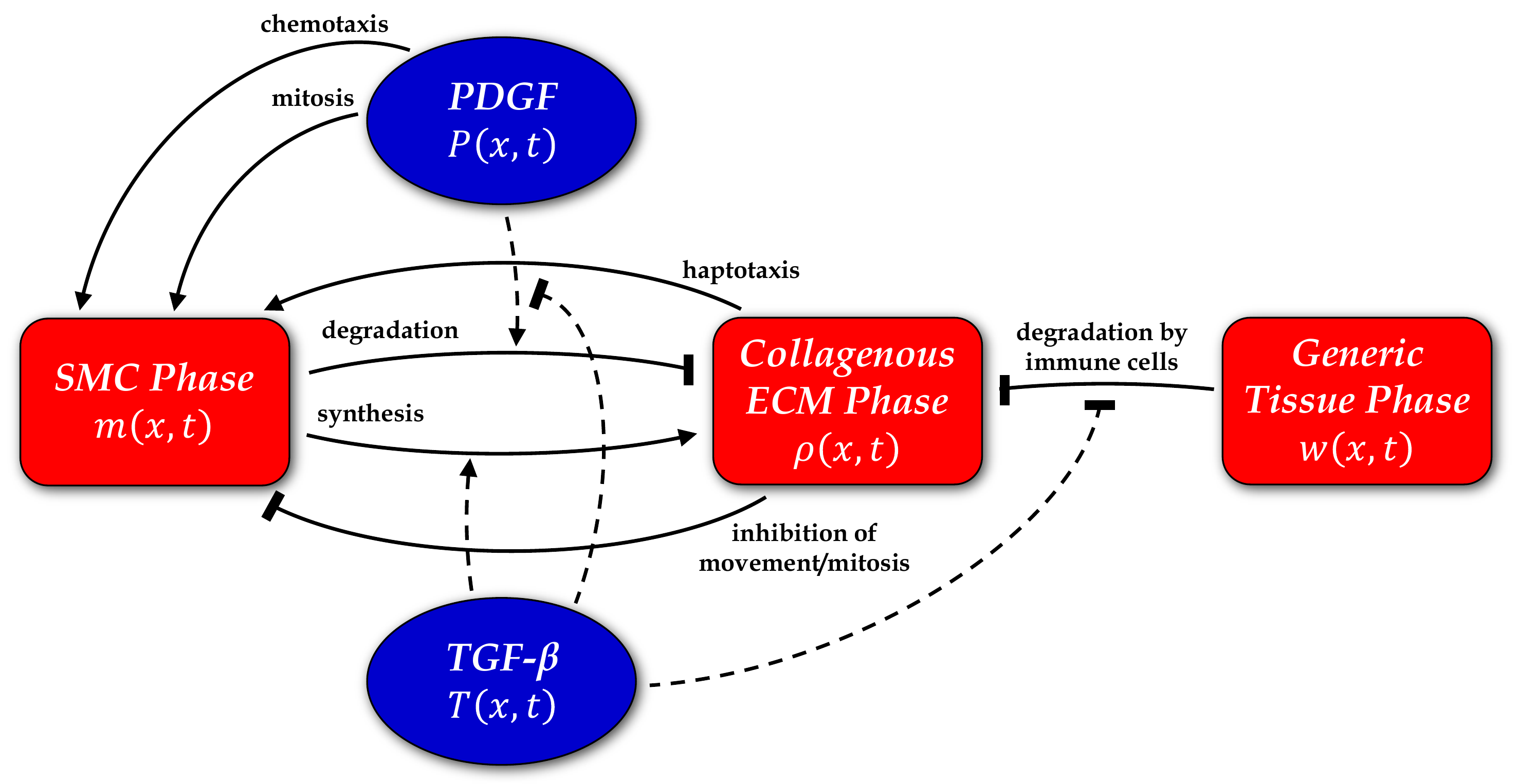}
		\caption{Schematic diagram of the primary interactions that regulate fibrous cap formation in the model. The volume-occupying phases are shown in red and the volumeless chemical growth factors are shown in blue. Labelled (solid) lines indicate interactions that act to increase (arrows) or decrease (bars) the SMC and ECM volume fractions in the cap region. Unlabelled (dashed) lines denote the stimulatory (arrows) and inhibitory (bars) effects of the growth factors on the synthesis and degradation of the ECM phase.}\label{ModSchem}
	\end{figure}

Summing the momentum balance equations~(\ref{mmom})--(\ref{wmom}), and substituting in the expressions~(\ref{taum})--(\ref{tauw}) gives the following equation for the tissue pressure gradient:
	\begin{equation} \label{summo}
		\frac{\partial p}{\partial x} = - \frac{\partial}{\partial x}\Big[ m \left( \Lambda + 2\rho\psi\right) \Big].
	\end{equation}
A second expression for the tissue pressure gradient is obtained by substituting the expressions~(\ref{Fmw}), (\ref{Frw}) and (\ref{tauw}) into equation~(\ref{wmom}) (note that equation~(\ref{Fmr}) becomes redundant at this juncture). Cancelling terms, we obtain:
	\begin{equation} \label{wmoms}
		\frac{\partial p}{\partial x} = k_{mw} m v_m - v_w \left(k_{mw} m + k_{w \rho} \rho \right).
	\end{equation}
Equating the right hand sides of equations~(\ref{summo}) and (\ref{wmoms}), and then using equation~(\ref{wvel}) to eliminate $v_w$, we derive the following expression for $v_m$ in terms of $m$, $\rho$ and $P$:
	\begin{equation} \label{mvel}
		v_m = - \left[ \, \frac{1 - m - \rho}{k_{mw} m \left(1 - \rho \right) + k_{w\rho} m \rho } \, \right] \frac{\partial}{\partial x} \Big[ m \left(\Lambda + 2\rho\psi \right) \Big].
	\end{equation}
In the interest of simplicity, we shall assume that the drag between the generic tissue phase and the SMC phase and between the generic tissue phase and the ECM phase is uniform (i.e. \ $k_{m w} = k_{w \rho} = k$). Making this simplification and substituting equation~(\ref{mvel}) back into equation~(\ref{mmas}), we arrive at the final mass balance relationship for the SMC phase in the intima:
	\begin{equation} \label{mmass}
		\frac{\partial m}{\partial t} = \frac{1}{k} \, \frac{\partial}{\partial x} \left[\, \left(1 - m - \rho \right) \frac{\partial}{\partial x} \Big[ m \left(\Lambda + 2\rho\psi \right) \Big] \, \right] + S_m\left(m,\rho,P\right).
	\end{equation}
Our reduced model therefore comprises equation~(\ref{mmass}) alongside the following expressions for $\rho$, $P$ and $T$:
	\begin{align}
		\frac{\partial \rho}{\partial t} &= S_\rho\left(m,\rho,P,T\right), \label{rmass}\\
		0 &= D_P \, \frac{\partial}{\partial x}\left[ \, \left(1 - m - \rho \right) \frac{\partial P}{\partial x} \, \right] + S_P\left(m,\rho,P\right), \label{cpmass} \\
		0 &= D_T \, \frac{\partial}{\partial x}\left[ \, \left(1 - m - \rho \right) \frac{\partial T}{\partial x} \, \right] + S_T\left(m,\rho,T\right). \label{ctmass}
	\end{align}
A schematic diagram that summarises the key model interactions involved in fibrous cap formation is provided in Figure~\ref{ModSchem}.

\subsection{Boundary and Initial Conditions} \label{ssBIC}
In this section we define the boundary conditions required to solve equations~(\ref{mmass}), (\ref{cpmass}) and (\ref{ctmass}), and the initial conditions required to solve equations~(\ref{mmass}) and (\ref{rmass}). Note that the assumptions that we make are largely consistent with those made previously in \citet{Wats18}.

Recall that in the model simplification in Section~\ref{ssMS} we have already assumed a zero-flux condition for SMCs at the endothelium. We therefore have the following boundary condition for the SMC phase:
	\begin{align}
		\frac{1}{k} \left(1 - m - \rho \right) \frac{\partial}{\partial x} \Big[m \left(\Lambda + 2\rho\psi \right) \Big] &= 0 \text{ at } x = 0 \nonumber \\
		\implies \frac{\partial} {\partial x} \Big[m \left(\Lambda + 2\rho\psi \right) \Big] &= 0 \text{ at } x = 0. \label{mBC0}
	\end{align}
Note that, in practice, this boundary condition stipulates that any flux of SMCs towards the endothelium at $x=0$ must be identically balanced by an equivalent flux of SMCs in the opposite direction.

SMC invasion of the plaque requires the activation of quiescent SMCs in the media by chemical signalling, and subsequent production of progeny that can negotiate the porous IEL \citep{Chap16, Jaco17}. The IEL provides a physical barrier to cell movement, so it is likely that a sustained and directed migratory response is required for synthetic SMCs to enter the plaque. We therefore assume that chemotaxis is the dominant mechanism of SMC migration across the IEL, and that any contribution from passive SMC diffusion or from ECM-mediated haptotaxis is negligible. We introduce the parameter $m_M<1$ to represent a constant volume fraction of activated medial SMCs, and assume a chemotactic flux of these cells into the intima in response to the PDGF gradient at the medial boundary:
	\begin{align}
		\frac{1}{k} \left(1 - m - \rho \right) \frac{\partial}{\partial x} \Big[m \left(\Lambda + 2\rho\psi \right) \Big] &= \frac{1}{k} \left(1 - m - \rho \right) \frac{d\Lambda}{d P} \, m_M \, \frac{\partial P}{\partial x} \text{ at } x = L \nonumber \\
		\implies \frac{\partial}{\partial x} \Big[m \left(\Lambda + 2\rho\psi \right) \Big] &= \frac{d\Lambda}{d P} \, m_M \, \frac{\partial P}{\partial x} \text{ at } x = L. \label{mBCL}
	\end{align}
There are several points to note regarding the format of this boundary condition. First, the model assumes that no SMCs will enter the intima in the absence of a PDGF gradient at the medial boundary. Second, the condition does not stipulate continuity of $m$ across the boundary (i.e.\ in general, $m \neq m_M$ at $x = L$). Finally, since we are assuming a closed system, the boundary condition implies that any flux of SMCs into the intima must be balanced by an equivalent efflux from the generic tissue phase into the media. Such an efflux of tissue constituents such as macrophages and foam cells may not be entirely physically realistic, but we resolve that our modelling assumptions remain reasonable provided that the total influx of plaque SMCs remains relatively small. An extended model that explicitly captures intimal growth could, of course, provide a more comprehensive treatment of the problem. However, we postpone modelling of domain growth for future work because we believe that the current approach provides significant insight into the cap formation process without the added complexity of tracking a moving boundary.

Stimulated endothelial cells and adherent platelets at the lesion site are believed to be the dominant sources of PDGF in the plaque \citep{Funa98}, and we model this influx of endothelium-derived PDGF with the following boundary condition:
	\begin{align}
		D_P \left(1 - m - \rho \right) \frac{\partial P}{\partial x} &= -\alpha_P \left(1 - m - \rho \right) \text{ at } x = 0 \nonumber \\
		\implies \frac{\partial P}{\partial x} &= -\frac{\alpha_P}{D_P} \text{ at } x = 0, \label{cPBC0}
	\end{align}
where $\alpha_P$ quantifies the rate of PDGF flux into the intima, and the term $\left(1-m-\rho\right)$ is included for consistency with the modulated PDGF transport inside the model domain. Note the corresponding implication that the absolute rate of PDGF influx into the plaque will reduce as the SMC and ECM phases accumulate at the luminal boundary. This introduces an implicit self-regulation mechanism in the model whereby SMCs can effectively downregulate the signal that recruits them to the cap region.

The boundary condition~(\ref{mBCL}) stipulates that SMC migration into the plaque from the media requires a negative PDGF gradient at the medial boundary. We therefore treat the IEL as a permeable membrane and assume that PDGF can diffuse into the media from the intima according to the following boundary condition:
	\begin{align}
		D_P \left(1 - m - \rho\right) \frac{\partial P}{\partial x} &= \sigma_P \left(P_M - P \right) \left(1 - m - \rho \right) \text{ at } x = L \nonumber \\ 
		\implies \frac{\partial P}{\partial x} &= \frac{\sigma_P}{D_P} \left(P_M - P \right) \text{ at } x = L. \label{cPBCL}
	\end{align}
Here, $\sigma_P$ denotes the permeability of the IEL to PDGF, $P_M$ represents the PDGF concentration in the media, and the term $\left(1-m-\rho\right)$ is again included for consistency with the modulated chemical transport inside the domain.

The boundary conditions that we impose for TGF-$\beta$ take an identical format to those defined for PDGF. Like PDGF, TGF-$\beta$ is also released by degranulating platelets and activated endothelial cells at the lesion site \citep{Toma12}. We therefore define the following condition at the endothelium:
	\begin{align}
		D_T \left(1 - m - \rho \right) \frac{\partial T}{\partial x} &= -\alpha_T \left(1 - m - \rho \right) \text{ at } x = 0 \nonumber \\
		\implies \frac{\partial T}{\partial x} &= -\frac{\alpha_T}{D_T} \text{ at } x = 0, \label{cTBC0}
	\end{align}
where $\alpha_T$ quantifies the rate of TGF-$\beta$ flux into the intima. At the medial boundary, we assume:
	\begin{align}
		D_T \left(1 - m - \rho\right) \frac{\partial T}{\partial x} &= \sigma_T \left(T_M - T \right) \left(1 - m - \rho \right) \text{ at } x = L \nonumber \\ 
		\implies \frac{\partial T}{\partial x} &= \frac{\sigma_T}{D_T} \left(T_M - T \right) \text{ at } x = L, \label{cTBCL}
	\end{align}
where $\sigma_T$ denotes the permeability of the IEL to TGF-$\beta$, and $T_M$ represents the TGF-$\beta$ concentration in the media. Note that the decision to allow TGF-$\beta$ to diffuse out of the domain is made simply for consistency with the boundary condition~(\ref{cPBCL}) for PDGF. The presence of an explicit TGF-$\beta$ gradient at the medial boundary is \emph{not} required to initiate cap formation in the model. 

Finally, we must define initial profiles for the SMC and collagenous ECM phases in the plaque. Observations of plaque development in mice indicate that the SMC population in the intima is negligible prior to cap formation \citep{Benn16}. The model equations cannot support a domain that is entirely devoid of SMCs (see the singularity that arises in equation~(\ref{mvel}), for example), so we initiate the plaque with a small and uniform initial SMC volume fraction:
	\begin{equation} \label{minit}
		m\left(x,0\right) = m_i, \text{ where } 0<m_i \ll 1.
	\end{equation}
We also assume a small and uniform initial profile for the collagenous ECM phase in the plaque:
	\begin{equation} \label{rinit}
		\rho\left(x,0\right) = \rho_i, \text{ where } 0<\rho_i \ll 1.
	\end{equation}

This initial collagen is assumed to represent a fraction of the ECM that exists in the intima prior to SMC invasion. The remaining constituents of the initial plaque ECM, whose quantities we do not explicitly track, are assumed to reside in the generic tissue phase and are notionally replaced by the deposition of new collagen over the course of a simulation.

\subsection{Model Non-Dimensionalisation}
Using tildes to denote dimensionless quantities, we rescale space $x$, time $t$, PDGF concentration $P$ and TGF-$\beta$ concentration $T$ as follows (note that the SMC volume fraction $m$ and the ECM volume fraction $\rho$ do not require rescaling):
	\begin{equation*}
		\widetilde{x} = \frac{x}{L}, \; \widetilde{t} = \frac{t}{t_0}, \; \widetilde{P} = \frac{P}{P_0}, \; \widetilde{T} = \frac{T}{T_0}, \; \widetilde{m} = m, \; \widetilde{\rho} = \rho,
	\end{equation*}
where $t_0$, $P_0$ and $T_0$ represent characteristic time, PDGF concentration and TGF-$\beta$ concentration values, respectively.  The model parameters can now be non-dimensionalised in the following way:
	\begin{equation*}
		\begin{gathered}
			\widetilde{\chi_P} = \frac{\chi_P t_0}{k L^2}, \; \widetilde{\kappa} = \kappa P_0, \; \widetilde{\chi_{\rho}} = \frac{2 \chi_{\rho} t_0}{k L^2}, \; \widetilde{\delta} = \frac{2 \delta t_0}{k L^2}, \; \widetilde{r_m} = r_m t_0, \\
			\widetilde{c_m} = \frac{c_m}{P_0}, \; \widetilde{\beta_m} = \beta_m t_0, \; 	\widetilde{r_s} = r_s t_0, \; \widetilde{c_s} = \frac{c_s}{T_0}, \; \widetilde{r_d} = r_d t_0, \\
			\widetilde{c_d} = \frac{c_d}{P_0}, \; \widetilde{\gamma_d} = \gamma_d T_0, \; 	\widetilde{\beta_{\rho}} = \beta_{\rho} t_0, \; \widetilde{\gamma_{\rho}} = \gamma_{\rho} T_0, \; \widetilde{\eta_P} = \frac{\eta_P L^2}{D_P}, \\
			\widetilde{\beta_P} = \frac{\beta_P L^2}{D_P}, \; \widetilde{\eta_T} = \frac{\eta_T L^2}{D_T}, \; \widetilde{\beta_T} = \frac{\beta_T L^2}{D_T}, \; 	\widetilde{\alpha_P} = \frac{\alpha_P L}{D_P P_0}, \\
			\widetilde{\alpha_T} = \frac{\alpha_T L}{D_T T_0}, \; \widetilde{\sigma_P} = \frac{\sigma_P L}{D_P}, \; \widetilde{\sigma_T} = \frac{\sigma_T L}{D_T}, \; \widetilde{P_M} = \frac{P_M}{P_0}, \; \widetilde{T_M} = \frac{T_M}{T_0}.
		\end{gathered}
	\end{equation*}
Dropping tildes for clarity, the corresponding dimensionless model equations, boundary conditions and initial conditions are:
\begin{gather}
	\begin{split} \label{mmasnd}
		\frac{\partial m}{\partial t} = \frac{\partial}{\partial x}{}& \left[ \, \left(1 - m - \rho \right) \frac{\partial}{\partial x} \Big[m \left(\Lambda + \rho \psi \right) \Big] \, \right] \\ &+ r_m m \left(1 - m - \rho \right) \left[ \, 1 + \frac{A_m P}{c_m + P} \, \right] - \beta_m m,	
	\end{split}\\
	\begin{split} \label{rmasnd}
		\frac{\partial \rho}{\partial t} = r_s m \left(1 - m - \rho \right) {}& \left[ \, 1 + \frac{A_s T}{c_s + T} \, \right] - r_d m \rho \left[ \, 1 + \frac{A_d P}{\left(c_d + P\right) \left(1 + \gamma_d T\right)} \, \right] \\ & - \beta_{\rho} \rho \left(1 - m - \rho \right) \left[ \, \frac{1+\varepsilon \, \gamma_{\rho} T}{1+\gamma_{\rho} T} \, \right],	
	\end{split}\\
	\frac{\partial}{\partial x}\left[\left(1 - m - \rho \right) \frac{\partial P}{\partial x}\right] = \eta_P m P \left(1 - m - \rho \right) + \beta_P P \left(1 - m - \rho \right), \label{cPmasnd} \\
	\frac{\partial}{\partial x} \left[\left(1 - m - \rho \right) \frac{\partial T}{\partial x} \right] = \eta_T m T \left(1 - m - \rho \right) + \beta_T T \left(1 - m - \rho \right), \label{cTmasnd} \\
	\frac{\partial} {\partial x} \Big[m \left(\Lambda + \rho\psi \right) \Big] = 0 \text{ at } x = 0, \;\; \frac{\partial} {\partial x} \Big[m \left(\Lambda + \rho\psi \right) \Big] = \frac{d\Lambda}{d P} \, m_M \, \frac{\partial P}{\partial x} \text{ at } x = L, \label{mBCnd} \\
	\frac{\partial P}{\partial x} = -\alpha_P \text{ at } x = 0, \;\; \frac{\partial P}{\partial x} = \sigma_P \left(P_M - P \right) \text{ at } x = L, \label{cPBCnd} \\
	\frac{\partial T}{\partial x} = -\alpha_T \text{ at } x = 0, \;\; \frac{\partial T}{\partial x} = \sigma_T \left(T_M - T \right) \text{ at } x = L, \label{cTBCnd} \\
	m\left(x,0\right) = m_i, \label{minitnd} \\
	\rho\left(x,0\right) = \rho_i, \label{rinitnd}
\end{gather}
wherein $\Lambda \! \left(P\right) = \dfrac{\chi_P}{1 + {\left(\kappa P \right)}^{n_P}}$ and $\psi \left(m,\rho\right) = - \chi_{\rho} + \dfrac{\delta m^{n_{\rho}}}{\left(1 - m - \rho\right)^{n_{\rho}}}$.

\subsection{Model Parameterisation} \label{ssParam}
A summary of the base case parameter values is provided in Table~\ref{params}. Our parameter selections have mostly been informed by data and observations from relevant \emph{in vitro} and \emph{in vivo} experimental studies. When appropriate data could not be found, we have chosen values that ensure biologically realistic results. Several important parameters (e.g.\ $A_m$, $c_m$, $\chi_P$, $\kappa$, $n_P$) have been assigned values that are dimensionally the same (or very similar) to those in our earlier model of cap formation. We provided detailed justifications for these parameter values in \citet{Wats18}, so below we focus on explaining the parameter selections that are unique to the current study.

	\begin{table}[hp]
		\centering
		\scriptsize
		\begin{tabular}{| l | p{8.5cm} | l | p{3cm} |}
			\hline
			Parameter & Description & Dimensionless & Reference \\
			& & value & \\ \hline
			$n_P$ & Exponent in SMC phase extra pressure function $\Lambda$ & 1.8 & \citet{Scha97} \\ \hline
			$\kappa$ & Reciprocal of reference PDGF concentration in SMC phase extra pressure function $\Lambda$ & 5.5 & \citet{Scha97} \\ \hline
			$\chi_P$ & Baseline SMC phase motility coefficient & 1.75 & \citet{Cai07} \\ \hline
			$n_\rho$ & Exponent in SMC phase extra pressure function $\psi$ & 2 & \\ \hline
			$\delta$ & SMC phase repulsion coefficient & 0.45 & \\ \hline
			$\chi_\rho$ & SMC phase affinity for ECM phase & 0.3 & \\ \hline
			$r_m$ & Baseline rate of SMC phase proliferation & 0.25 & \citet{Bret86} \\
			& & & \citet{Chap16} \\ \hline
			$A_m$ & Maximal factor of PDGF-stimulated increase in rate of SMC proliferation & 14 & \citet{Munr94} \\ \hline
			$c_m$ & PDGF concentration for half-maximal increase in rate of SMC proliferation & 1.5 & \citet{Munr94} \\ \hline
			$\beta_m$ & Rate of SMC phase loss & 0.6 & \citet{Lutg99} \\ \hline
			$r_s$ & Baseline rate of ECM phase synthesis by SMCs & 1.8 & \citet{Reif12} \\ \hline
			$A_s$ & Maximal factor of TGF-$\beta$-stimulated increase in rate of ECM synthesis & 1 & \citet{Kubo03} \\ \hline
			$c_s$ & TGF-$\beta$ concentration for half-maximal increase in rate of ECM synthesis & 0.3 & \citet{Kubo03} \\ \hline
			$r_d$ & Baseline rate of ECM phase degradation by SMCs & 1.5 & \\ \hline
			$A_d$ & Maximal factor of PDGF-stimulated increase in rate of ECM degradation by SMCs & 4 & \citet{Borr06} \\ \hline
			$c_d$ & PDGF concentration for half-maximal increase in rate of ECM degradation by SMCs & 2.5 & \citet{Borr06} \\ \hline
			$\gamma_d$ & Reciprocal of reference TGF-$\beta$ concentration for inhibition of PDGF-stimulated ECM degradation by SMCs & 0.5 & \citet{Borr06} \\ \hline
			$\beta_\rho$ & Baseline rate of ECM phase degradation by immune cells & 0.75 & \\ \hline
			$\varepsilon$ & Maximal factor of TGF-$\beta$-stimulated decrease in rate of ECM degradation by immune cells & 0.25 & \citet{Vada01} \\ \hline
			$\gamma_\rho$ & Reciprocal of reference TGF-$\beta$ concentration for inhibition of ECM degradation by immune cells & 10 & \citet{Vada01} \\ \hline
			$\eta_P$ & Rate of PDGF uptake by SMC phase & 2.5 & \\ \hline
			$\beta_P$ & Rate of PDGF decay & 0.2 & \citet{Haug06} \\ \hline
			$\eta_T$ & Rate of TGF-$\beta$ uptake by SMC phase & 2.5 & \\ \hline
			$\beta_T$ & Rate of TGF-$\beta$ decay & 20 & \citet{Wake90} \\ \hline
			$m_M$ & Volume fraction of synthetic SMCs in media & 0.01 & \\ \hline
			$\alpha_P$ & Rate of PDGF influx from endothelium & 0.7 & \\ \hline
			$\sigma_P$ & Permeability of IEL to PDGF & 4 & \\ \hline
			$P_M$ & PDGF concentration in media & 0 & \\ \hline
			$\alpha_T$ & Rate of TGF-$\beta$ influx from endothelium & 2.5 & \\ \hline
			$\sigma_T$ & Permeability of IEL to TGF-$\beta$ & 4 & \\ \hline
			$T_M$ & TGF-$\beta$ concentration in media & 0 & \\ \hline
			$m_i$ & Initial SMC volume fraction in intima & $10^{-4}$ & \citet{Benn16} \\ \hline
			$\rho_i$ & Initial ECM volume fraction in intima & 0.02 & \citet{Reif12} \\ \hline
		\end{tabular}
		\caption{Base case parameter values. The final column reports any references that have been used to calculate individual parameter values. Values that do not have references have been chosen to ensure biologically reasonable results. Unless otherwise stated, all reported simulations use these values.} \label{params}
	\end{table}
	
\emph{In vivo} studies in the ApoE-deficient mouse have shown that plaque fibrous caps typically form over the course of several months \citep{Koza02}. We therefore assume a characteristic timescale of approximately 1 month and set $t_0 = 2.5 \times 10^6$~\si{\second}. For the domain length, we set $L = 120$~\si{\micro\metre}. This value is based on data from \citet{Reif12}, who measured intimal thickness in the diseased aortic arches of mice around the time of initial plaque SMC infiltration. For PDGF and TGF-$\beta$, we assume the characteristic concentrations $P_0=10$~\si{\ng\per\ml} and $T_0=1$~\si{\ng\per\ml}. These values reflect reported concentrations of PDGF in platelets (15--50~\si{\ng\per\ml}; \citet{Huan88}) and TGF-$\beta$ in blood plasma (2--12~\si{\ng\per\ml}; \citet{Wake95}).

In a detailed model of the fibroblast response to PDGF in dermal wound healing, \citet{Haug06} estimated PDGF diffusion coefficient and decay rate values that suggest a PDGF diffusion distance of approximately 300~\si{\micro\metre}. Based on these estimates, we set our dimensionless PDGF decay rate to be $\beta_P=0.2$. As TGF-$\beta$ has an almost identical molecular weight to PDGF, we expect that TGF-$\beta$ will diffuse at a similar rate to PDGF in the plaque tissue. However, active TGF-$\beta$ has been reported to decay around two orders of magnitude faster than PDGF \citep{Wake90}, and we therefore set $\beta_T=20$. The other parameters that determine the plaque growth factor profiles are more difficult to estimate. We assume, for consistency, that $\sigma_P=\sigma_T$ and we choose corresponding values for $\alpha_P$ and $\alpha_T$ that give reasonable growth factor concentration ranges in the intima. The response of the model to changes in the values of $\alpha_P$ and $\alpha_T$ will be an important focus of our numerical simulation studies in Section \ref{ssNumSim}. Note that we also set $P_M=T_M=0$ because neither growth factor is known to have a prominent source in the medial tissue.    

ECM remodelling in the model is governed by a range of concentration-dependent stimulatory and inhibitory effects of PDGF and TGF-$\beta$. The parameter values in these relationships have all been informed by data from \emph{in vitro} experiments. \citet{Kubo03} used SMCs derived from human atherosclerotic plaques to show that TGF-$\beta$ can increase baseline collagen synthesis up to 2-fold, with a half-maximal response occurring for a TGF-$\beta$ concentration in the interval 0.2--0.5~\si{\ng\per\ml}. We therefore set $A_s=1$ and $c_s=0.35$. Values for $\varepsilon$ and $\gamma_\rho$ are based on results from \citet{Vada01}, who studied the impact of TGF-$\beta$ on tumour necrosis factor (TNF)-$\alpha$-induced expression of MMP-9 in monocytes. Note that TNF-$\alpha$ is known to be an important inflammatory cytokine in atherosclerosis progression \citep{Ursc15}. \citet{Vada01} showed that, in the presence of 1~\si{\ng\per\ml} TNF-$\alpha$, MMP-9 activity decreased with increasing TGF-$\beta$ concentration and was reduced by around 50--75\% for TGF-$\beta$ concentrations in the range 0.1--1~\si{\ng\per\ml}. We therefore estimate $\varepsilon=0.25$ and $\gamma_\rho=10$. The parameter values that quantify the competing effects of PDGF and TGF-$\beta$ on ECM remodelling by SMCs have been calculated using data from \citet{Borr06}. \citet{Borr06} showed that SMCs demonstrate very similar qualitative patterns of MMP-2 and MMP-9 release in the presence of varying concentrations of PDGF and TGF-$\beta$. We therefore base our quantitative estimates on the averaged responses of these two MMPs to PDGF and TGF-$\beta$. In the absence of TGF-$\beta$, the data suggest an average 2.5-fold increase in basal MMP release at 20~\si{\ng\per\ml} PDGF and an average 4-fold increase at 50~\si{\ng\per\ml} PDGF. When 5~\si{\ng\per\ml} TGF-$\beta$ is introduced at 50~\si{\ng\per\ml} PDGF, the release of MMPs is cut, on average, by around 50\%. Based on these observations, we select $A_d=4$, $c_d=2.5$ and $\gamma_d=0.5$.  

Of course, the parameters discussed above determine only how ECM remodelling is modified in the presence of growth factors. The underlying baseline rates of ECM synthesis and degradation ($r_s$, $r_d$ and $\beta_\rho$) must be determined separately. It is difficult to ascertain typical values for these parameters during \emph{in vivo} plaque progression, so we base our parameter selections on the following considerations. First, we assume that, as invasive plaque SMCs undertake a process of continual ECM remodelling, the values of $r_s$ and $r_d$ should be of a similar order. Furthermore, we assume that the value of $r_s$ must be sufficiently large to allow plaque collagen to accumulate on a timescale similar to that reported in \citet{Reif12}. Second, we assume that the value of $\beta_\rho$ should be smaller than the value of $r_d$. We make this assumption because, while inflammatory cells such as macrophages may degrade the plaque ECM more aggressively than SMCs, these cells make up only a fraction of the content of the generic tissue phase. Based on these assumptions, we estimate $r_s=1.8$, $r_d=1.5$ and $\beta_\rho=0.75$.  

In \citet{Wats18}, we assumed a dimensionless baseline SMC proliferation rate $r_m=0.02$. This value was estimated based on previous experimental observations that suggested a proliferative index of around 1\% for plaque SMCs (i.e.\ 1\% of plaque SMCs displaying evidence of mitotic activity at a given point in time). However, as discussed in Section~\ref{sIntro}, recent SMC lineage tracing studies have indicated that plaque SMC proliferation may be more prominent than previously appreciated. For example, \citet{Chap16} have recently reported a plaque SMC proliferative index of approximately 4\%. Furthermore, this value was determined at an advanced stage of plaque progression, where a significant SMC population had already assembled. Hence, accounting for likely effects such as cell-cell and cell-ECM contact inhibition, we anticipate that the plaque SMC proliferative index could be even larger in early cap formation. We therefore assume a significant increase in our previous estimate and set $r_m=0.25$. To our knowledge, no recent lineage tracing study has reported an equivalent calculation for the SMC apoptotic index. Hence, for the SMC death rate in the model, we use the apoptotic index of 1\% that was calculated by \citet{Lutg99} and set $\beta_m=0.6$. This value corresponds to a typical SMC apoptosis time of approximately 8 hours.

Finally, we briefly comment on the rationale for our parameterisation of the SMC affinity for the ECM phase $\chi_\rho$. Expansion of the flux term in equation (\ref{mmasnd}) shows that $\chi_\rho$ represents an effective haptotaxis coefficient for SMC movement up gradients in the collagen-rich ECM. The extent to which haptotactic SMC migration contributes to \emph{in vivo} cap formation is not well known. We therefore make a conservative initial estimate of $\chi_\rho=0.3$ and will investigate the sensitivity of the model to changes in this value in Section \ref{ssNumSim}. Note that this conservative estimate for $\chi_\rho$ also arises due to practical considerations. Numerical simulations indicate that the model can become ill-posed if the value of $\chi_\rho$ is chosen to be too large. This issue appears to be related to the onset of \emph{negative} SMC diffusion, which is an occurrence that we envisage to be unlikely during \emph{in vivo} plaque growth. We therefore avoid such parameter regimes in the simulations that are presented in this paper. In Section \ref{sDiscuss}, we will discuss this issue in greater detail and propose amendments that can be made to the model to reduce or eliminate the ill-posed region of the parameter space.

\section{Results} \label{sResults}
Before presenting numerical solutions of the model system (\ref{mmasnd})--(\ref{rinitnd}), we begin this section by performing a simplified steady state analysis that provides useful insight into the relationship between the SMC phase and the collagenous ECM phase in the model plaque.

\subsection{Analytical Results} \label{ssAnalyt}
The mathematical model derived in Section~\ref{sMFP} provides a significant challenge for analytical studies. Indeed, given the highly nonlinear nature of the model equations, it seems very unlikely that closed-form expressions could be derived for any spatially inhomogeneous steady state solutions. Worthwhile progress can be made, however, if we exploit the absence of a flux term in the ECM equation and investigate the steady state ECM phase solution at a fixed spatial position for fixed values of the other model variables. This analysis, which we will use to interpret the numerical results in Section~\ref{ssNumSim}, allows us to understand why a particular combination of steady state values for $P$, $T$ and $m$ correspond to a particular steady state value for $\rho$ at a given location in the plaque.  

Consider a fixed point inside the model domain, and let us assume that, at that point, the PDGF and TGF-$\beta$ concentrations take the fixed (steady state) values $P^*$ and $T^*$. Assuming that the SMC volume fraction has the fixed (steady state) value $m^*$ at the same point, equation (\ref{rmasnd}) gives the following expression for the steady state ECM volume fraction $\rho^*$:
	\begin{equation} \label{ECMSS}
		R_s m^* \left(1-m^*-\rho^*\right) - R_d m^* \rho^* - B_{\rho} \rho^* \left(1-m^*-\rho^*\right) = 0,
	\end{equation}
where $\rho^* + m^* < 1$. Note that, in the above expression, we have introduced the positive constants $R_s\left(T^*\right)=r_s \left[ 1 + \frac{A_s T^*}{c_s + T^*} \right]$, $R_d\left(P^*,T^*\right)=r_d \left[ 1 + \frac{A_d P^*}{\left(c_d + P^*\right) \left(1 + \gamma_d T^*\right)} \right]$, and $B_{\rho}\left(T^*\right)=\beta_{\rho} \left[ \frac{1+\varepsilon \, \gamma_{\rho} T^*}{1+\gamma_{\rho} T^*} \right]$. These constants represent the (local) steady state rates of ECM synthesis by SMCs, ECM degradation by SMCs, and ECM degradation by immune cells, respectively. Rearranging equation~(\ref{ECMSS}) gives the following quadratic in $\rho^*$:
	\begin{equation} \label{ECMquad}
		{\rho^*}^2 + \left[m^*\left(1-\frac{R_s}{B_{\rho}}-\frac{R_d}{B_{\rho}}\right)-1\right] \rho^* + \frac{R_s}{B_{\rho}} \, m^* \left(1-m^*\right) = 0.
	\end{equation}
Setting $\mu\left(T^*\right)=\frac{R_s}{B_{\rho}}$ and $\lambda\left(P^*,T^*\right)=\frac{R_d}{B_{\rho}}$, this quadratic gives two possible solutions $\rho{^*_{\pm}}$ for the steady state ECM volume fraction in terms of $m^*$:
	\begin{equation} \label{ECMquadsols}
		\rho{^*_{\pm}}\left(m^*\right) = \frac{1}{2} \left[ \, 1 - m^* \left(1 - \mu - \lambda \right) \, \pm \, \sqrt{ {\Big[ 1 - m^* \left(1 - \mu - \lambda \right) \Big]}^2 - 4 \mu m^* \left( 1 - m^*\right) } \, \right],
	\end{equation}
where $0<m^*<1$. It is trivial to show that the discriminant of equation (\ref{ECMquadsols}) is strictly positive and, hence, that the two solutions are real-valued, positive and distinct. However, for either solution to be physically meaningful, it must satisfy the condition $0<\rho{^*_{\pm}}<1-m^*$. It can therefore be shown that $\rho{^*_-}$ gives the only admissable steady state solution and, furthermore, that this steady state is always stable. From here, we drop the subscript on $\rho{^*_-}$ and use $\rho^*$ to refer to the negative root in equation~(\ref{ECMquadsols}). 

For fixed values of $\mu$ and $\lambda$, plots of $\rho^*\left(m^*\right)$ for $m\in\left(0,1\right)$ demonstrate a biphasic behaviour where $\rho^*$ initially increases from zero with increasing $m^*$ but then diminishes back towards zero as $m^*$ approaches one. The region of decreasing $\rho^*$ is attributable to the fact that as $m^*$ increases, ECM synthesis is increasingly inhibited by the reduced availability of generic tissue material. An interesting consequence of this biphasic relationship between $\rho^*$ and $m^*$ is that, for each pair of growth factor concentrations $P^*$ and $T^*$ (or parameter values $\mu$ and $\lambda$), there exists a SMC volume fraction $\widehat{m}^*$ that produces a \emph{maximum} steady state ECM volume fraction $\widehat{\rho}\,^*$. Therefore, in what follows, we shall derive expressions for both $\widehat{m}^*$ and $\widehat{\rho}\,^*$, and use these expressions to explore the conditions that lead to optimal ECM deposition at a given point in the plaque.

We begin by looking for stationary points of the function $\rho^*\left(m^*\right)$, whose first derivative is given by the following:
	\begin{equation} \label{dSSECM}
		\frac{d\rho^*}{dm^*} = \frac{1}{2} \left[ \mu + \lambda - 1 + \frac{1 + \mu - \lambda - m^*\Big[ {\left(1 - \mu - \lambda\right)}^2 + 4\mu\Big]}{\sqrt{ {\Big[ 1 - m^* \left(1 - \mu - \lambda \right) \Big]}^2 - 4 \mu m^* \left( 1 - m^*\right) }} \, \right].
	\end{equation}
Before proceeding, we remark that the second derivative of $\rho^*\left(m^*\right)$ is strictly negative and, hence, that any stationary point found from equation (\ref{dSSECM}) will indeed be a maximum. Setting equation (\ref{dSSECM}) equal to zero, rearranging terms and squaring leads to the following quadratic equation in $\widehat{m}^*$:
	\begin{equation} \label{mmaxquad}
		\Big[{\left(1-\mu-\lambda\right)}^2+4\mu\Big] {{}\widehat{m}^*}^2 - 2\left(1 + \mu - \lambda \right) \widehat{m}^* + 1 - \lambda = 0,
	\end{equation}
which has corresponding solutions:
	\begin{equation} \label{mmaxquadsols}
		\widehat{m}_\pm^* = \frac{1 + \mu - \lambda \, \pm \, \sqrt{\lambda}\left(1 - \mu - \lambda\right)}{{\left(1- \mu - \lambda \right)}^2 + 4 \mu}.
	\end{equation}
The physically meaningful solution, which must satisfy the condition $0<\widehat{m}_\pm^*<1$, is again given by the negative root. Selecting the negative root and dropping the subscript, it is possible to show that $\widehat{m}^*$ can be written in the reduced form:
	\begin{equation} \label{optmmula}
		\widehat{m}^*\left(\mu,\lambda\right) = \frac{1+\sqrt{\lambda}}{1+\mu+\lambda+2\sqrt{\lambda}},
	\end{equation}
which, upon substitution into the negative root in equation (\ref{ECMquadsols}), gives the corresponding form of $\widehat{\rho}\,^*$:
	\begin{equation} \label{optrmula}
		\widehat{\rho}\,^*\left(\mu,\lambda\right) = \frac{\mu}{1+\mu+\lambda+2\sqrt{\lambda}}.
	\end{equation}

The dependence of $\widehat{\rho}\,^*$ and $\widehat{m}^*$ on $\mu$ and $\lambda$ can be inspected visually by plotting 2D heatmaps of the right hand sides of equations~(\ref{optmmula}) and (\ref{optrmula}). However, these plots have limited utility for the interpretation of numerical simulation results because, at steady state in a given simulation, the values of $\mu$ and $\lambda$ at each location in the plaque are not immediately obvious. This is because $\mu$ and $\lambda$ depend not only on the local growth factor concentrations $P^*$ and $T^*$, but also on the ten parameters ($r_s$, $A_s$, $c_s$, $r_d$, $A_d$, $c_d$, $\gamma_d$, $\beta_{\rho}$, $\varepsilon$, $\gamma_{\rho}$)  that appear in the definitions of $R_s$, $R_d$ and $B_{\rho}$. Therefore, to support the interpretation of the numerical results that will be presented in Section~\ref{ssNumSim}, we instead plot heatmaps of $\widehat{\rho}\,^*$ and $\widehat{m}^*$ as functions of $P^*$ and $T^*$ where the above parameters have been assigned their base case values (Figure~\ref{heat(P,T)}). Note that expressions for $\widehat{\rho}\,^*\left(P^*,T^*\right)$ and $\widehat{m}^*\left(P^*,T^*\right)$ can be written down explicitly (see the relevant definitions above), but we omit these here for brevity.

	\begin{figure}
		\centering
		\begin{subfigure}[b]{0.49\textwidth}
			\includegraphics[width=\textwidth]{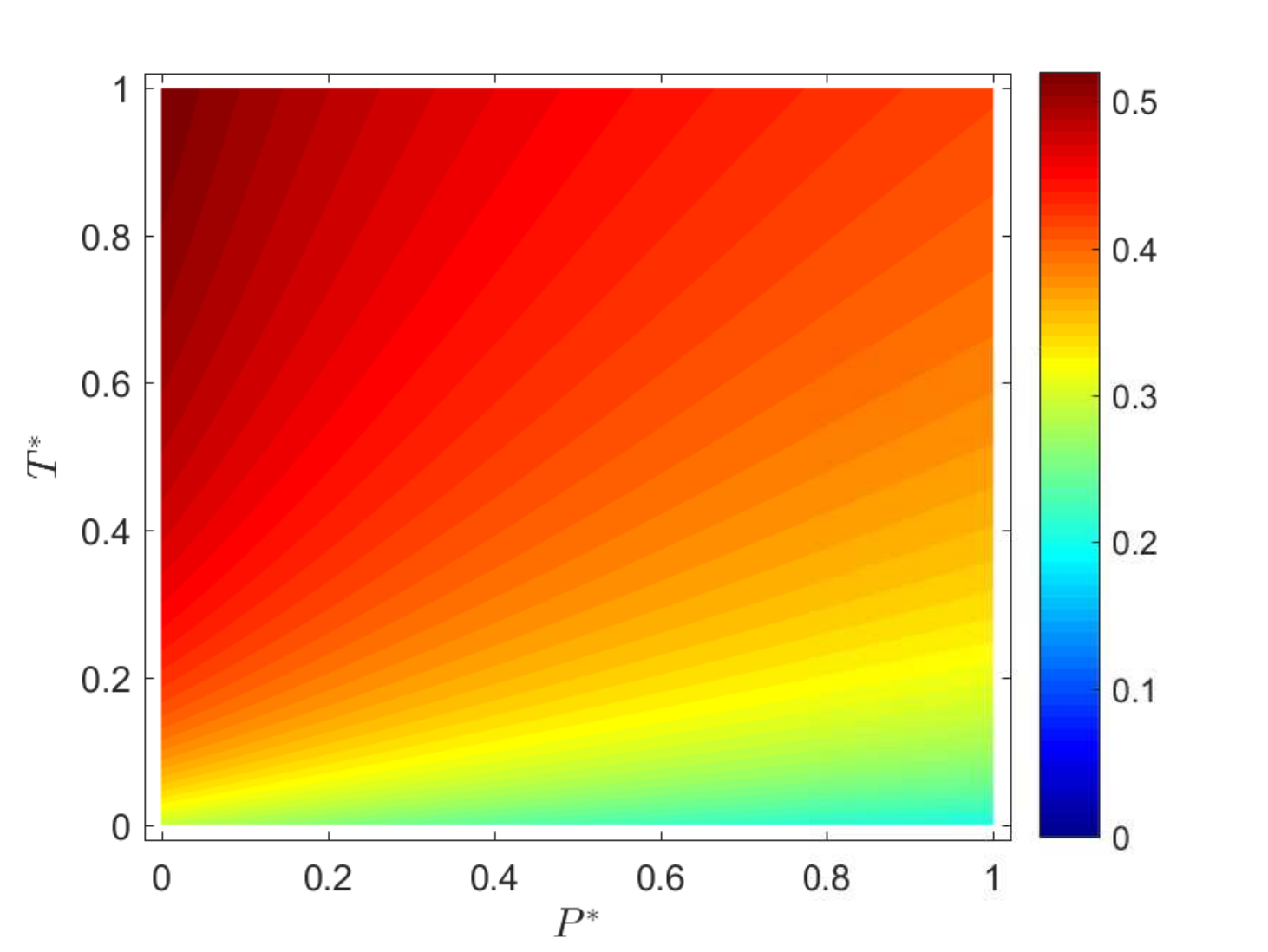}
			\caption{} \label{heatr(P,T)}
		\end{subfigure}
		\begin{subfigure}[b]{0.49\textwidth}
			\includegraphics[width=\textwidth]{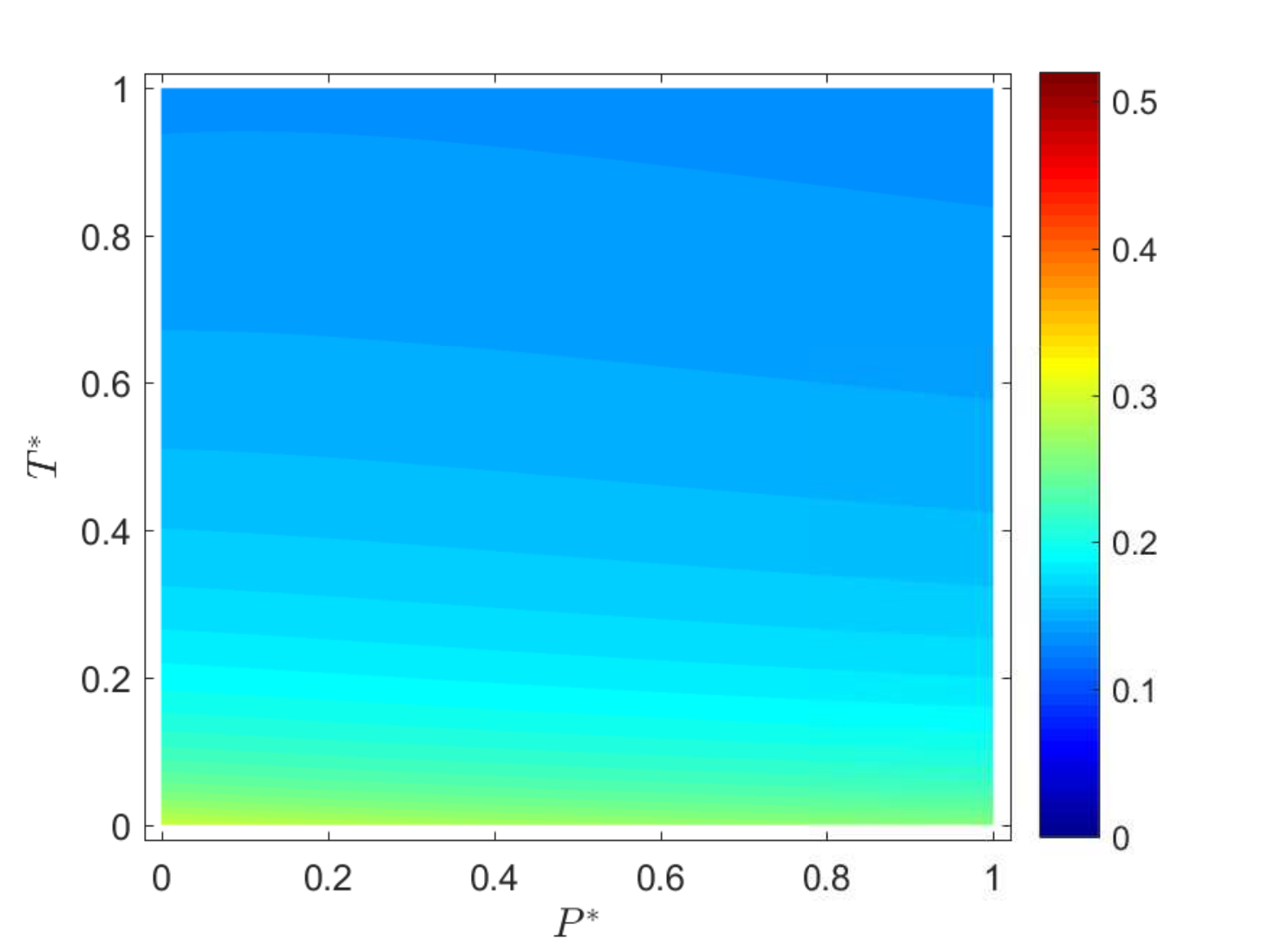}
			\caption{} \label{heatm(P,T)}
		\end{subfigure}
		\caption{Heatmaps that show how (a) the maximum steady state ECM volume fraction $\widehat{\rho}\,^*$ and (b) the corresponding SMC volume fraction $\widehat{m}^*$ at a fixed point in the plaque vary with the local growth factor concentrations $P^*$ and $T^*$ for $\left(P^*,T^*\right) \in [0,1] \times [0,1]$. Note that the parameters that appear in the definitions of $\widehat{\rho}\,^*$ and $\widehat{m}^*$ ($r_s$, $A_s$, $c_s$, $r_d$, $A_d$, $c_d$, $\gamma_d$, $\beta_{\rho}$, $\varepsilon$, $\gamma_{\rho}$) have been assigned their base case values (see Table~\ref{params}). Explicit expressions for $\widehat{\rho}\,^*\left(P^*,T^*\right)$ and $\widehat{m}^*\left(P^*,T^*\right)$ are not provided but can be inferred from the expressions in equations (\ref{optmmula}) and (\ref{optrmula}), and the definitions in the text after equations (\ref{ECMSS}) and (\ref{ECMquad}).} \label{heat(P,T)}
	\end{figure}

Figures~\ref{heatr(P,T)} and \ref{heatm(P,T)} present respective plots of $\widehat{\rho}\,^*\left(P^*,T^*\right)$ and $\widehat{m}^*\left(P^*,T^*\right)$ for $\left(P^*,T^*\right) \in [0,1] \times [0,1]$. These intervals for $P^*$ and $T^*$ reflect the typical ranges of dimensionless growth factor concentrations that will be considered in our numerical simulations in Section~\ref{ssNumSim}. Recall that, for each parameter combination $\left(P^*,T^*\right)$, these heatmaps indicate the maximum steady state ECM volume fraction that could be reached at a fixed point in the plaque (Figure~\ref{heatr(P,T)}), and the local SMC volume fraction that is required to achieve this maximum (Figure~\ref{heatm(P,T)}). For SMC volume fractions that are \emph{either above or below} those plotted for each $\left(P^*,T^*\right)$ in Figure~\ref{heatm(P,T)}, the corresponding steady state ECM volume fractions will be \emph{smaller} than those plotted in Figure~\ref{heatr(P,T)}. Intuitively, $\widehat{\rho}\,^*$ takes its largest values when $T^*$ is large and $P^*$ is small (i.e.\ high net rate of ECM synthesis and low net rate of ECM degradation), and its smallest values when $T^*$ is small and $P^*$ is large (i.e.\ low net rate of ECM synthesis and high net rate of ECM degradation). The corresponding behaviour of $\widehat{m}^*$, on the other hand, is less intuitive. Figure~\ref{heatm(P,T)} indicates that, over the majority of the plotted $\left(P^*,T^*\right)$ parameter space, the maximum ECM volume fraction can be generated by a relatively small volume fraction (0.1--0.2) of SMCs. 

Figure~\ref{heatr(P,T)} indicates that, while both growth factors can have a significant impact on plaque ECM levels, $\widehat{\rho}\,^*$ is much more sensitive to $T^*$ than to $P^*$ over the relevant concentration range. This is because TGF-$\beta$ plays a role not just in synthesising the collagenous ECM, but also in preventing its degradation. To quantify the relative sensitivity to $T^*$ and $P^*$, we note that, in general, an increase in $T^*$ from zero to one causes $\widehat{\rho}\,^*$ to increase by approximately 0.2, while a similar change in $P^*$ causes $\widehat{\rho}\,^*$ to decrease by approximately 0.1. Of course, it is also important to consider the SMC levels that are required to attain these maximum ECM volume fractions at different plaque growth factor levels. For $T^*$ close to zero, Figure~\ref{heatm(P,T)} indicates that an SMC volume fraction of around 25--30\% is required to maximise the steady state ECM volume fraction at around 20--30\%. However, for $T^*$ close to one, a maximum steady state ECM volume fraction of 40--50\% can be generated by an SMC volume fraction of less than 15\%. The model therefore suggests that, in the presence of favourable TGF-$\beta$ levels, it is possible to synthesise around twice as much collagenous ECM with about half as many tissue-synthesising SMCs.  

The above results provide a useful tool with which to interpret the final outcome of cap formation in a given model simulation. However, it is also important to emphasise that this simplified analysis does not allow us to readily predict cap formation efficacy in a particular case. We have focussed on a fixed spatial position in the plaque and have disregarded dynamic interactions between the various model variables, including the role of PDGF in SMC recruitment and the mechanical feedback between the SMCs and the ECM that they deposit. In the next section, we investigate the importance of these factors, and others, by performing numerical simulations with the full system of model equations.

\subsection{Numerical Results} \label{ssNumSim}
We begin this section by presenting results from a simulation with the base case parameter values. This is followed by a sensitivity analysis that investigates the impact of key parameters on cap formation. All numerical solutions have been obtained by discretising equations~(\ref{mmasnd})--(\ref{cTBCnd}) in space and/or time using a central differencing scheme for $P$ and $T$, the trapezoidal rule for $\rho$, and the Crank-Nicolson method for $m$. The resulting systems of nonlinear equations were solved sequentially for each variable at each timestep by iteration. 

\subsubsection{Base Case Simulation} \label{sssBase}
\begin{figure}
		\centering		
  		\includegraphics[height=5.8cm]{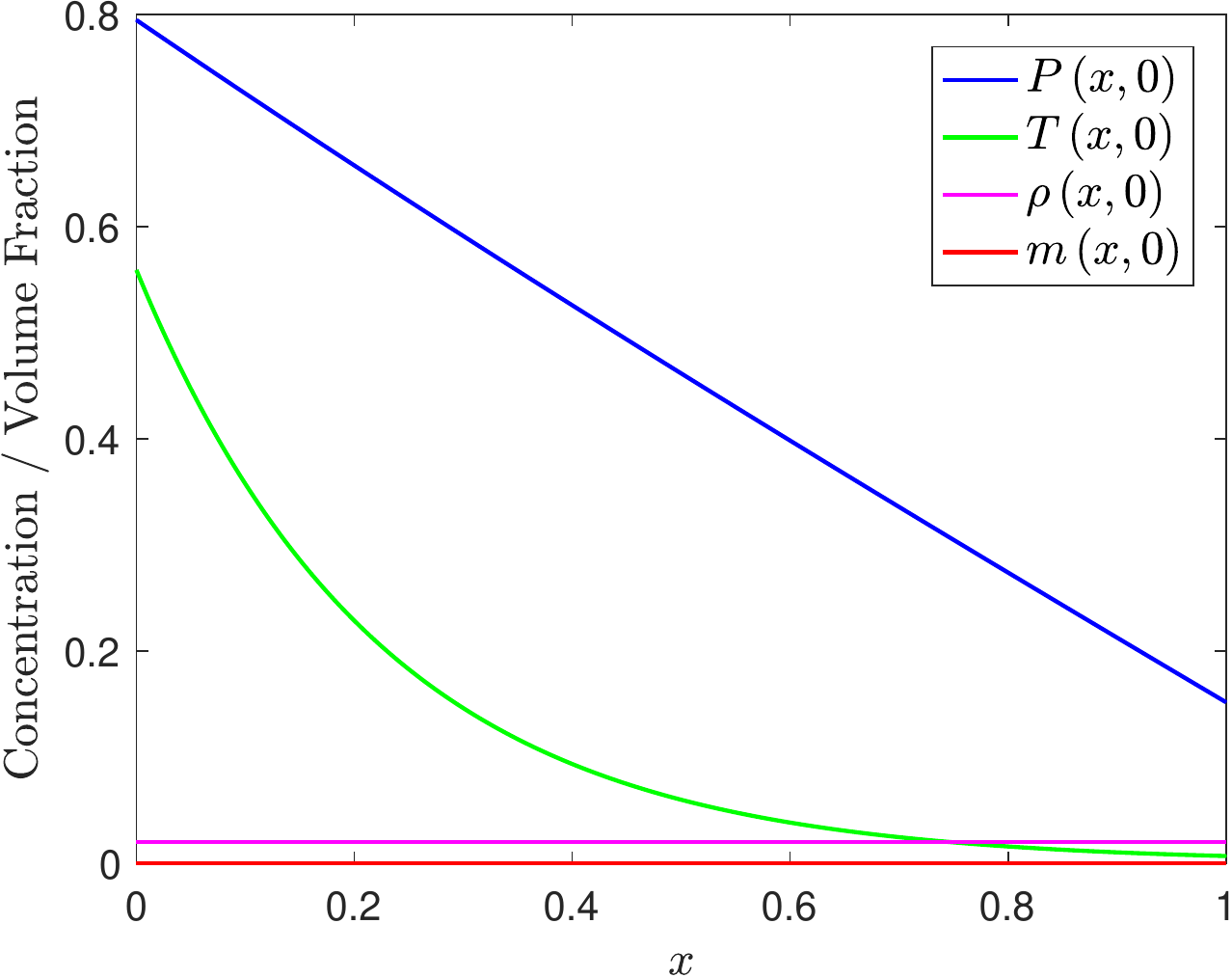}
		\caption{Initial conditions for each of the model variables in the base case simulation. The PDGF concentration is shown in blue, the TGF-$\beta$ concentration in green, the ECM volume fraction in magenta, and the SMC volume fraction in red.} \label{BaseInit}
	\end{figure}

We initiate the base case simulation by prescribing small and spatially uniform volume fractions for $m$ and $\rho$ in the plaque. The initial profiles are presented in Figure~\ref{BaseInit} alongside approximate initial solutions for the plaque growth factor concentrations $P$ and $T$. By using the fact that $1-m-\rho=\text{const}$ and $m\approx0$ at $t=0$, the $P$ and $T$ profiles represent exact solutions of simplified forms of equations~(\ref{cPmasnd}) and (\ref{cTmasnd}) (i.e.\ $\frac{\partial^2 G}{\partial x^2}=\beta_G G\text{, for } G=P, T$), subject to the corresponding boundary conditions~(\ref{cPBCnd}) and (\ref{cTBCnd}). An important feature to note about the initial growth factor profiles is the distinct difference between the distributions of endothelium-derived PDGF and TGF-$\beta$. The PDGF profile, which is diffusion-dominated ($\beta_P \ll 1$), shows an approximately linear decrease across the intima, while TGF-$\beta$, which is degradation-dominated ($\beta_T \gg 1$), shows a rapid exponential decline. Consequently, TGF-$\beta$ is more strongly localised to the endothelium and, of course, to the desired region of collagenous cap formation.

\begin{figure}
		\centering
		\begin{subfigure}[b]{0.45\textwidth}
			\includegraphics[width=\textwidth]{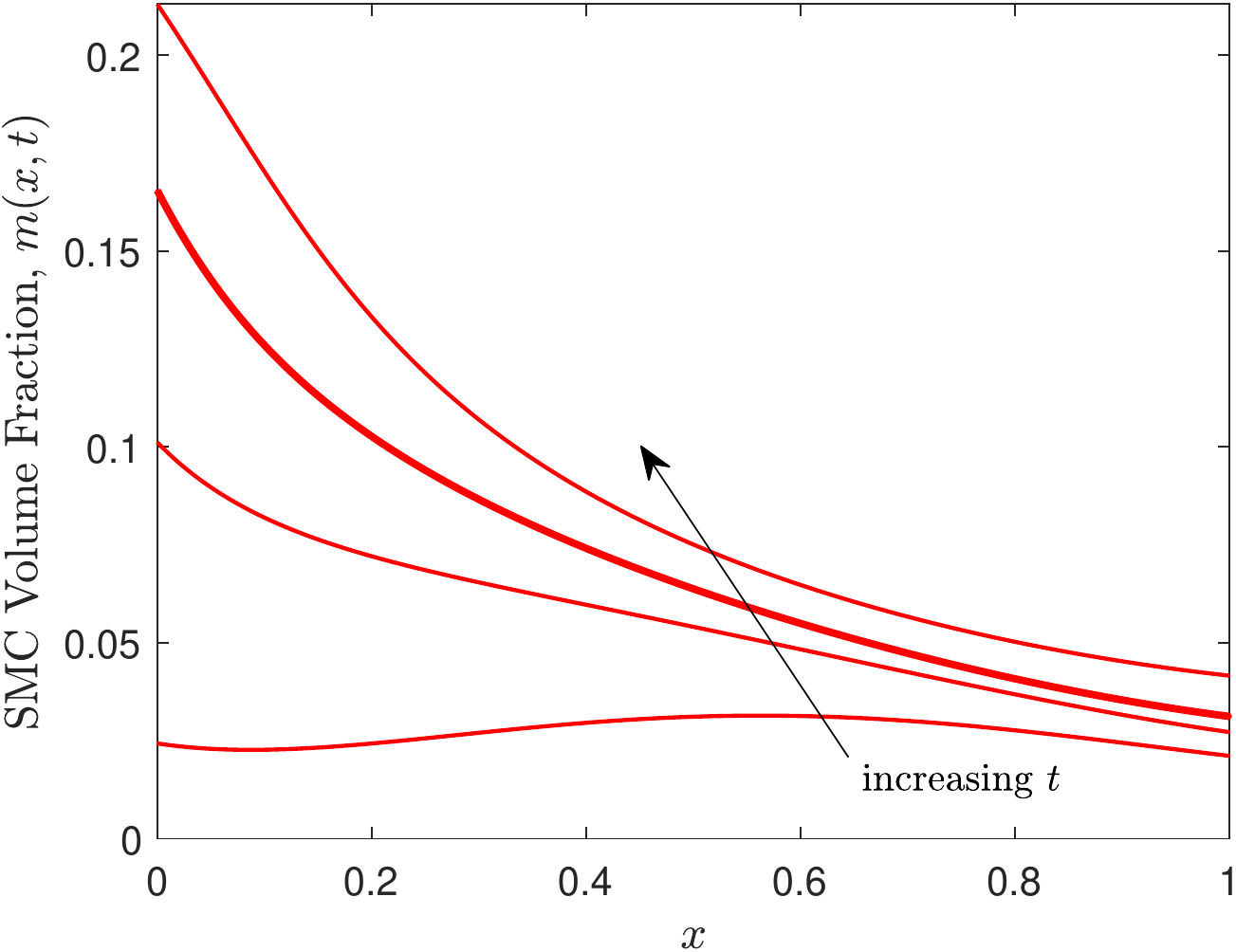}
			\caption{} \label{mBase}
		\end{subfigure}
		\hspace{0.05\textwidth}
		\begin{subfigure}[b]{0.45\textwidth}
			\includegraphics[width=\textwidth]{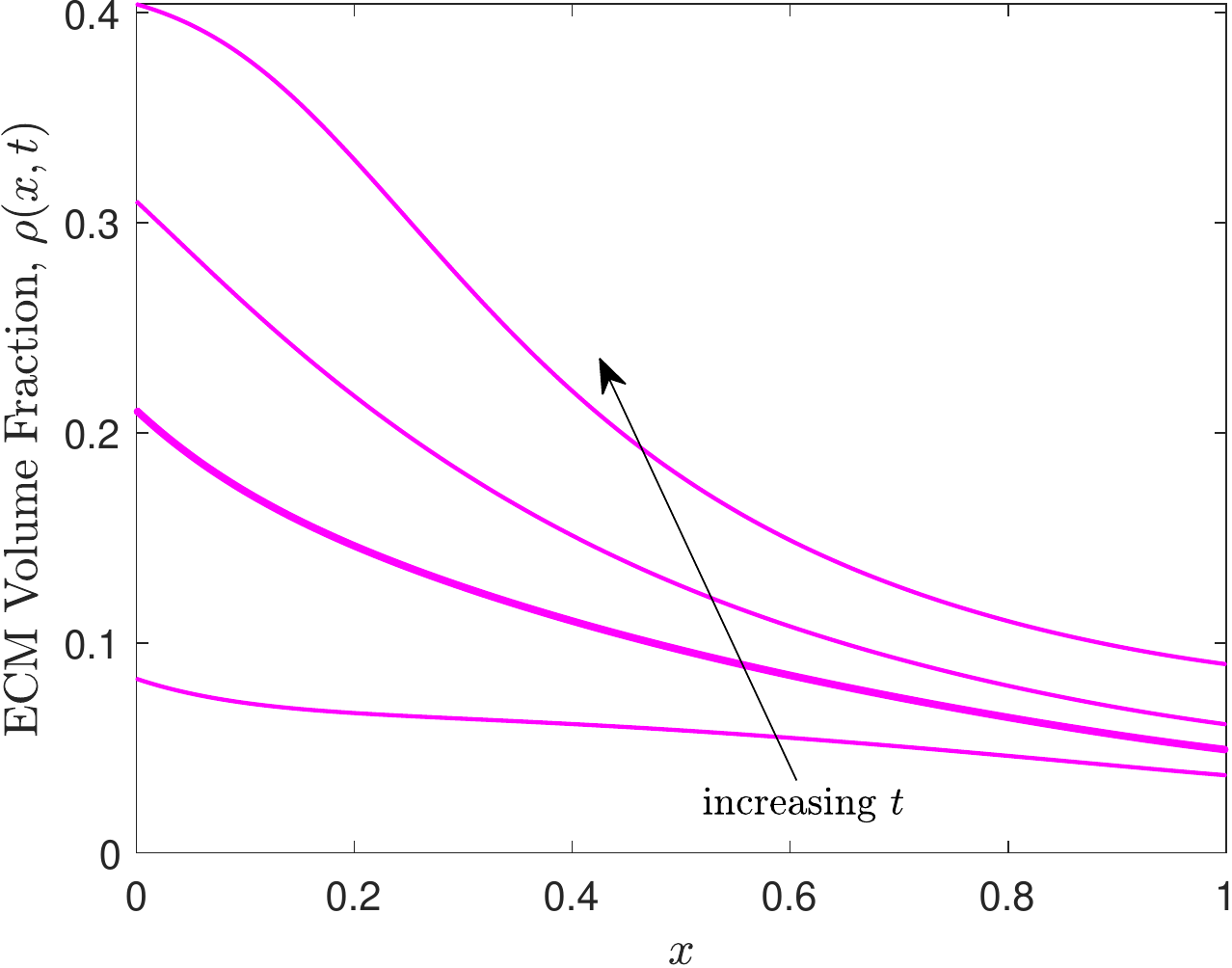}
			\caption{} \label{rBase}
		\end{subfigure} \\
		\begin{subfigure}[b]{0.45\textwidth}
			\includegraphics[width=\textwidth]{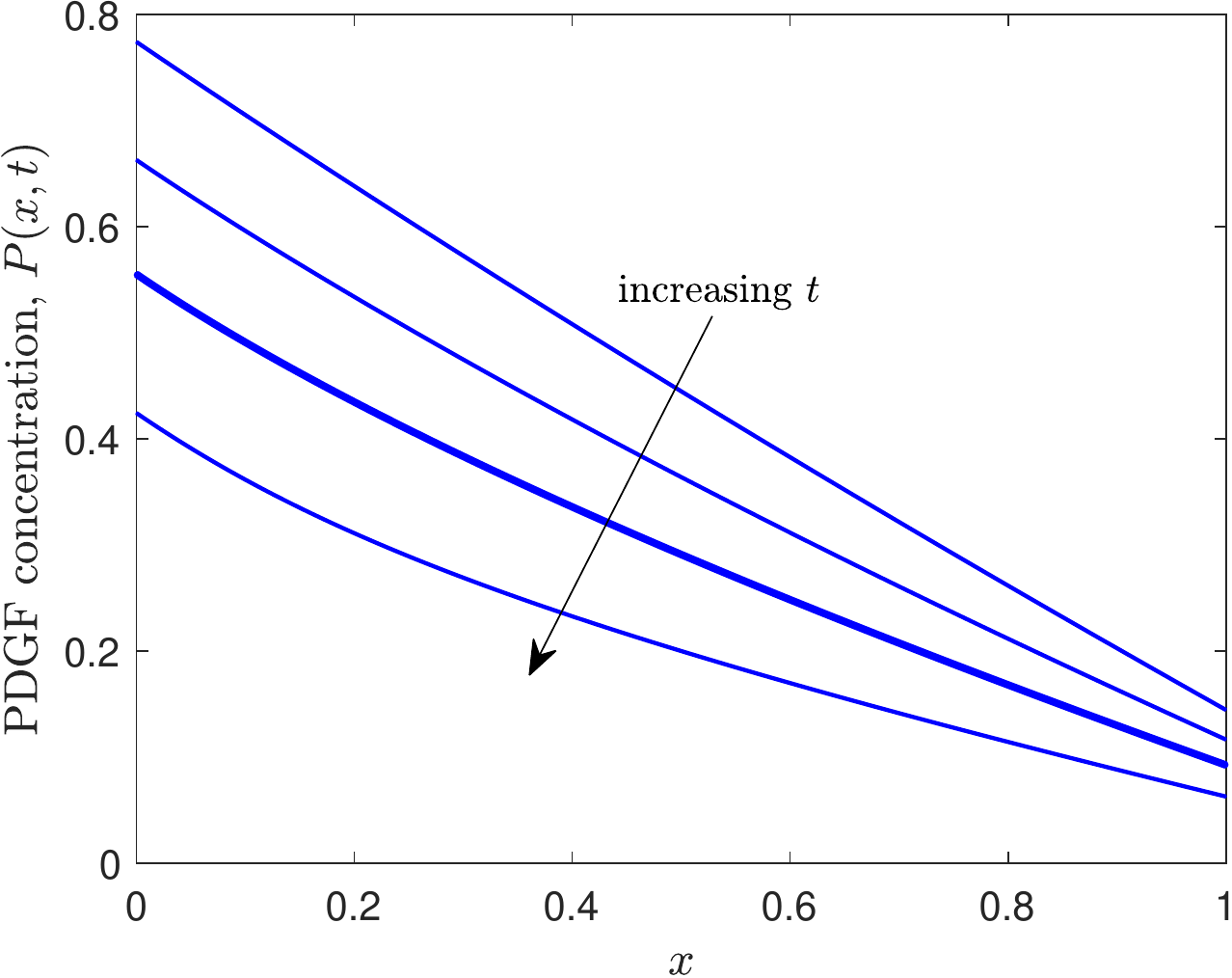}
			\caption{} \label{PBase}
		\end{subfigure}
		\hspace{0.05\textwidth}
		\begin{subfigure}[b]{0.45\textwidth}
			\includegraphics[width=\textwidth]{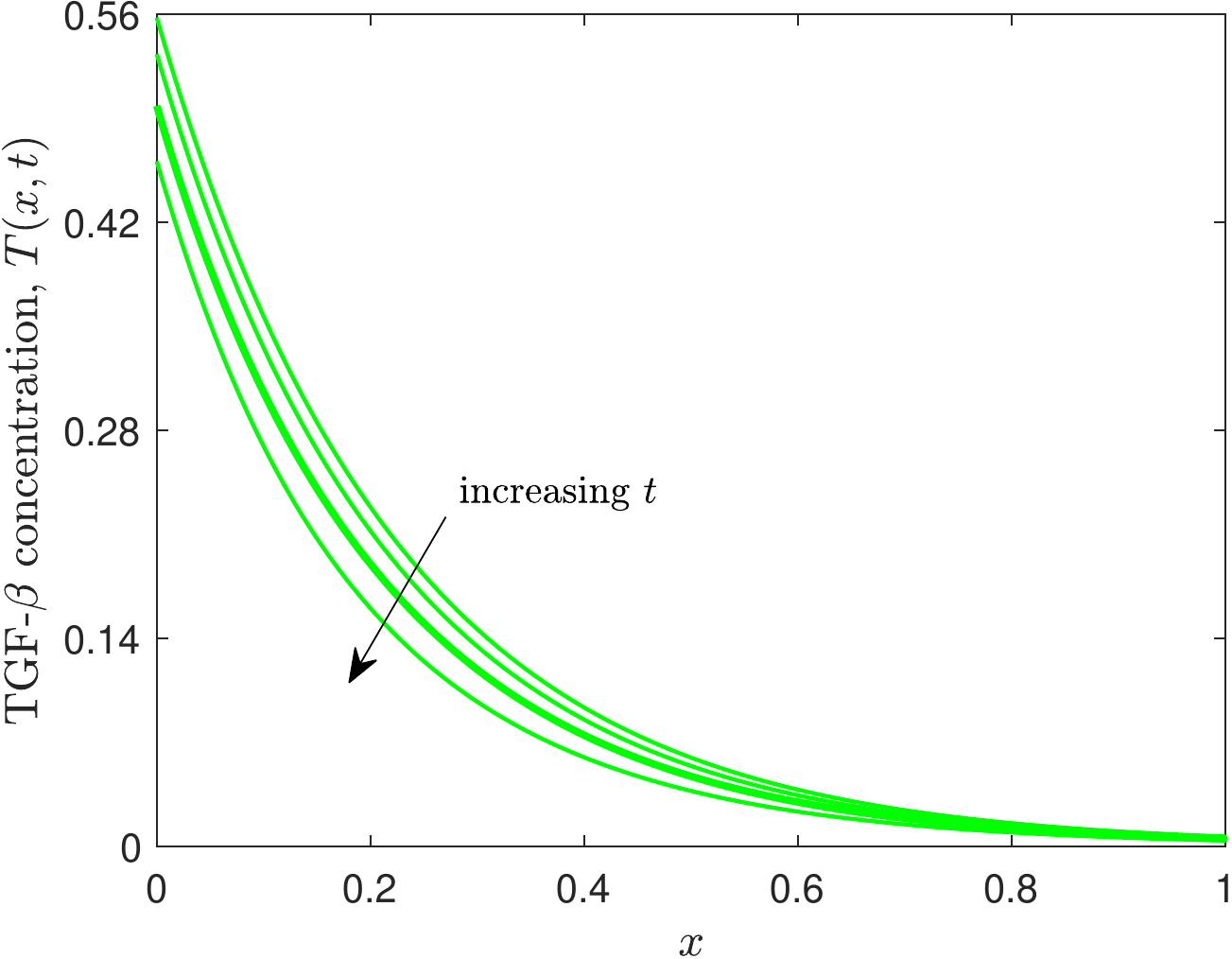}
			\caption{} \label{TBase}
		\end{subfigure}
		\caption{(a) SMC volume fraction, (b) ECM volume fraction, (c) PDGF concentration and (d) TGF-$\beta$ concentration profiles in the plaque at several times during the base case simulation. Arrows indicate the direction of sequential time points, which correspond to $t=1.3,2.0,2.7$ and 6.0 for the ECM phase and $t=0.8,1.5,2.0$ and 4.0 for the other variables. Data at the time point $t=2.0$ has been emboldened on each plot.} \label{mrPTBase}
	\end{figure}

The time evolution of the model variables in the base case simulation is presented in Figure~\ref{mrPTBase} (note that the time points for the ECM phase plots in Figure~\ref{rBase} do not correspond exactly to those for the plots of the other variables). The results demonstrate significant accumulations of SMCs (Figure~\ref{mBase}) and ECM in the cap region over the course of the simulation and a concurrent reduction in the overall concentrations of PDGF (Figure~\ref{PBase}) and TGF-$\beta$ (Figure~\ref{TBase}). The reduction in plaque PDGF and TGF-$\beta$ levels is partly due to uptake by the invading SMCs, but can mostly be attributed to reduced chemical diffusion through the increasingly populated cap region. Note that the PDGF profile undergoes a more dramatic change than the TGF-$\beta$ profile as a consequence of this modulated diffusivity.

Figure~\ref{mBase} shows that the media-derived model SMCs respond to the diffusible PDGF signal by migrating towards the endothelium and proliferating rapidly over the first 2 months of simulation time. Due to the ongoing increase in ECM deposition and concurrent reduction in PDGF levels, the SMCs divide more slowly after this time point and the plaque SMC population peaks at around 4 months. Interestingly, Figure~\ref{rBase} demonstrates that cap formation occurs on a longer timescale than that required for the ECM-synthesising SMCs to accumulate. After 2 months of simulation time, only 50\% of the final ECM content has been deposited, and the approximate steady state ECM profile is not attained until after about 6 months. The results in Figures~\ref{mBase} and \ref{rBase} are qualitatively and quantitatively consistent with the experimental observations of \citet{Reif12}. For ApoE knockout mice fed a high-fat diet for 24 weeks, \citet{Reif12} found that plaque SMC content increased from around 3\% at 8 weeks to around 7\% at both 16 and 24 weeks, while plaque collagen content increased from around 3\% at 8 weeks to around 17\% at 16 weeks and then to around 24\% at 24 weeks. These values show evidence of an early saturation in SMC numbers and a delayed saturation in collagen levels, both of which occur on similar timescales to those depicted in our simulation data.

\begin{figure}
		\centering
		\includegraphics[height=5.8cm]{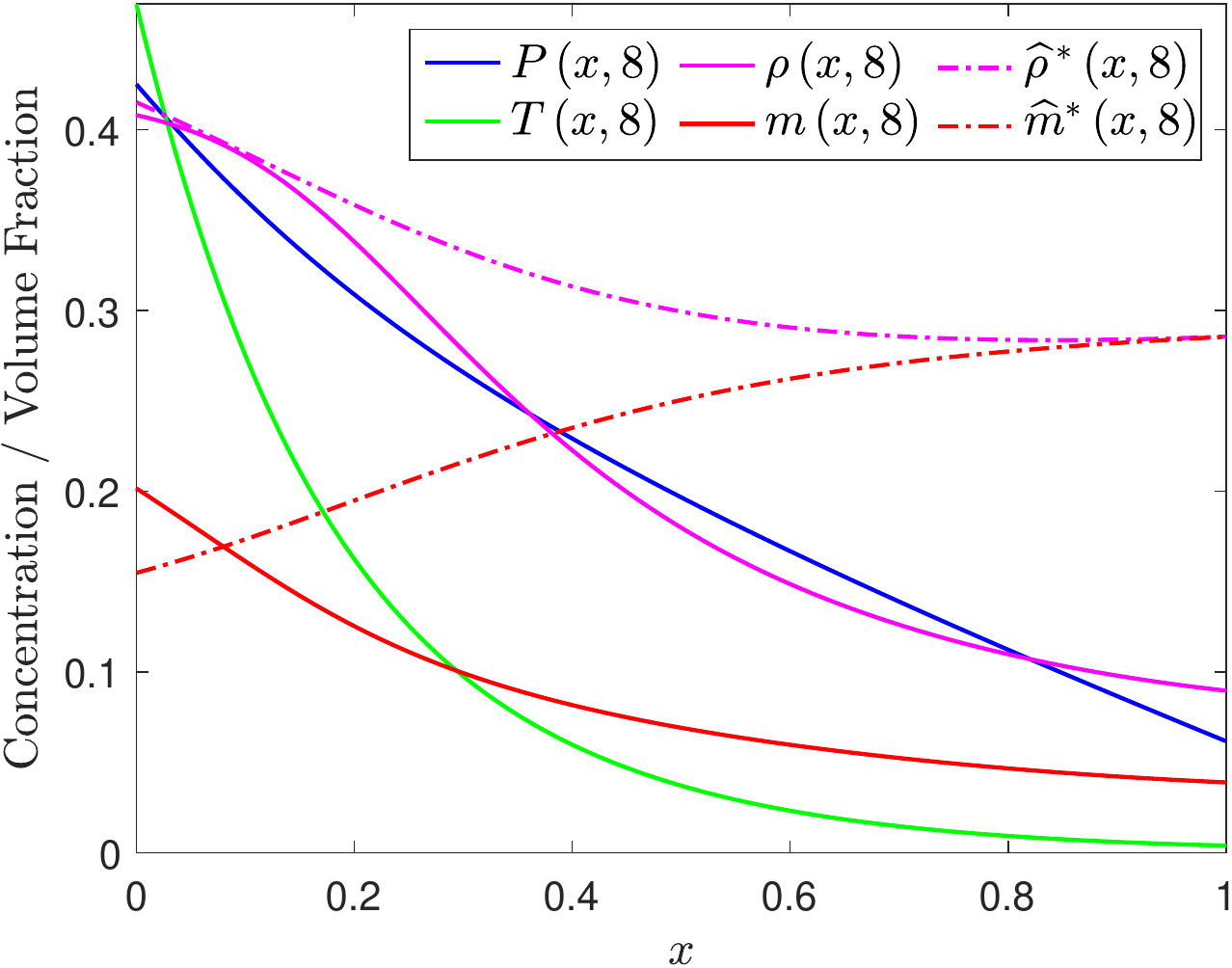}
		\caption{Base case simulation results that show approximate steady state profiles (solid lines) for the PDGF concentration (blue), TGF-$\beta$ concentration (green), ECM volume fraction (magenta) and SMC volume fraction (red). Dotted lines represent the analytically-derived maximum steady state ECM volume fraction $\widehat{\rho}\,^*$ (pink) and corresponding steady state SMC volume fraction $\widehat{m}^*$ (red) at each $x$. The $\widehat{m}^*$ and $\widehat{\rho}\,^*$ values are functions of the local PDGF and TGF-$\beta$ concentrations and have been calculated at each $x$ by equations~(\ref{optmmula}) and (\ref{optrmula}), respectively. All plots taken at time $t=8$.}\label{BaseFin}
	\end{figure}

Figure~\ref{BaseFin} presents the profile of each model variable in the plaque at the later time $t=8$ (solid lines). The simulation is close to steady state at this time point, and the only notable change from the results at the latest time points in Figure~\ref{BaseInit} is a small reduction in the SMC volume fraction throughout the domain. This reduction in SMC numbers appears to be due to the delayed accumulation of ECM, which causes increased inhibition of SMC migration and SMC proliferation at a late stage of cap formation. At $t=8$, the total volume fractions of SMCs and collagenous ECM in the plaque are approximately 8.6\% and 21.4\%, respectively. These values are consistent with observations from several studies of plaque growth in the ApoE knockout mouse \citep{Mall01, Clar06, Reif12}.

In Figure~\ref{BaseFin}, we have also plotted the values of $\widehat{m}^*$ and $\widehat{\rho}\,^*$ at each spatial position in the plaque (dotted lines). These plots, which have been constructed using equations~(\ref{optmmula}) and (\ref{optrmula}), represent the maximum ECM volume fraction ($\widehat{\rho}\,^*$) that could be attained at each position $x$, and the corresponding SMC volume fraction ($\widehat{m}^*$) that would be required to achieve this maximum. For each $x$, the values of $\widehat{m}^*$ and $\widehat{\rho}\,^*$ are given by a unique pair of values $\mu$ and $\lambda$, which depend on the local PDGF and TGF-$\beta$ concentration values $P\left(x,8\right)$ and $T\left(x,8\right)$. The $\widehat{m}^*$ and $\widehat{\rho}\,^*$ curves show that, in the majority of the plaque tissue ($x\gtrsim0.08$), the simulated SMC phase volume fraction $m\left(x,8\right)$ is below the level required to maximise ECM deposition. For $x\lesssim0.08$, the extent of ECM deposition is again sub-optimal, but, in this case, the reduced ECM volume fraction $\rho\left(x,8\right)$ is the result of \textit{over}-recruitment of SMCs. Interestingly, Figure~\ref{BaseFin} shows that the simulated ECM volume fraction remains relatively close to maximal levels throughout the region proximal to the endothelium (i.e.\ $x\lesssim0.2$). This is despite the fact that the simulated SMC volume fraction is considerably larger than $\widehat{m}^*$ for $x\approx0$ and considerably smaller than $\widehat{m}^*$ for $x\approx0.2$.

\begin{figure}
		\centering
		\includegraphics[height=5.8cm]{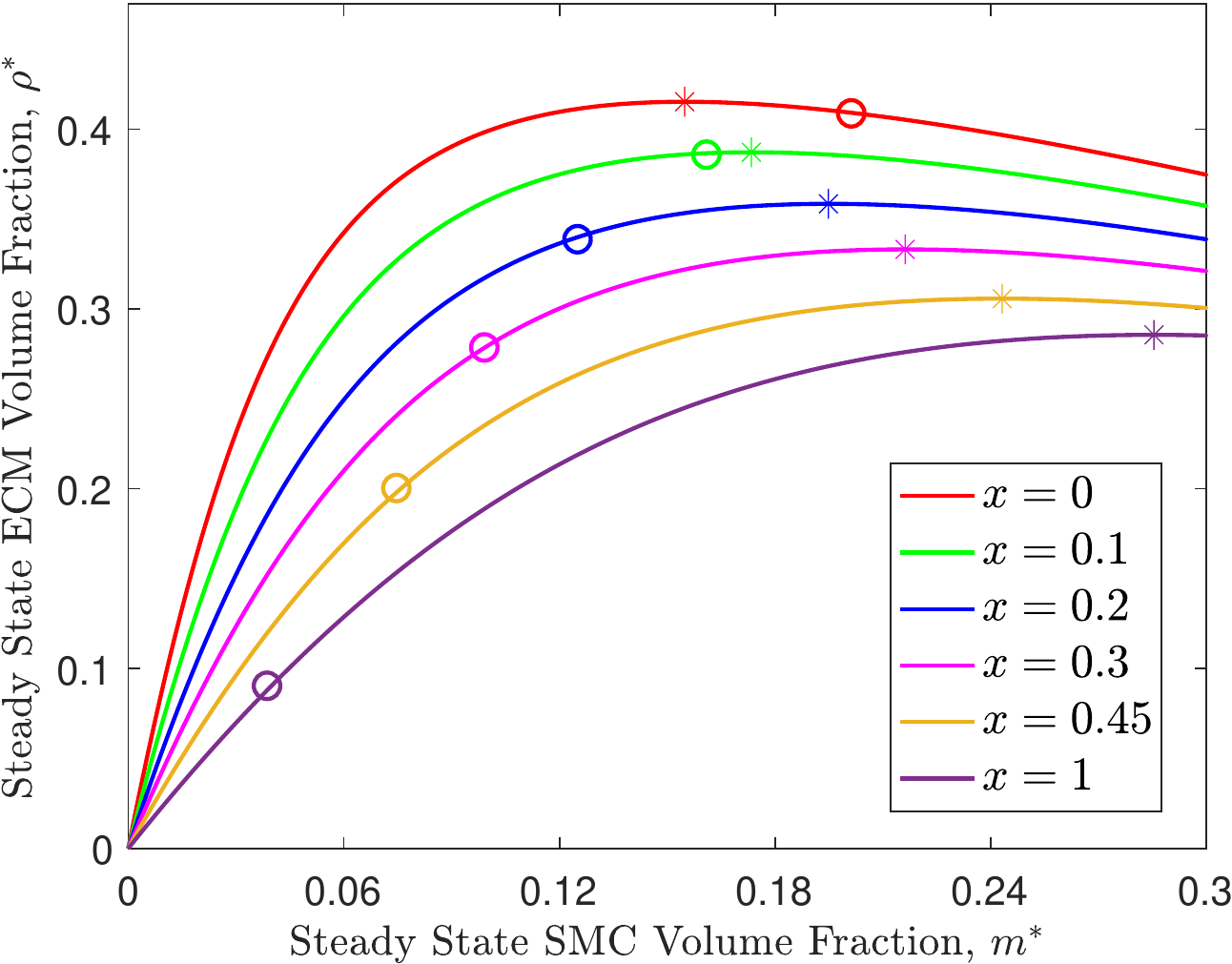}
		\caption{Plot that compares near-steady state $m$ and $\rho$ values for several values of $x$ in the base case simulation with the corresponding analytical steady state results derived in Section~\ref{ssAnalyt}. Solid lines represent the expressions $\rho^*\left(m^*\right)$ (negative root in equation~(\ref{ECMquadsols})) at each of the spatial positions $x=0$ (red), $x=0.1$ (green), $x=0.2$ (blue), $x=0.3$ (magenta), $x=0.45$ (orange) and $x=1$ (purple). Each $\rho^*\left(m^*\right)$ curve is a function of the local PDGF and TGF-$\beta$ concentrations. Individual data points represent the near-steady state $\left(m,\rho\right)$ values from the base case simulation (open circles) and the maximum $\left(\widehat{m}^*,\widehat{\rho}\,^*\right)$ of each individual $\rho^*\left(m^*\right)$ curve (asterisks) at each $x$. All curves and data points taken at time $t=8$.}\label{analytic}
	\end{figure}

Figure~\ref{analytic} presents additional data that helps to explain why \textit{near-maximal} ECM deposition can be attained for a wide range of SMC volume fractions near the endothelium in the base case simulation. For a series of spatial positions in the model plaque ($x=0, 0.1, 0.2, 0.3, 0.45, 1$), Figure~\ref{analytic} plots the analytical expression for the steady state ECM volume fraction $\rho^*\left(m^*\right)$ (negative root in equation~(\ref{ECMquadsols}); solid lines). Each individual $\rho^*\left(m^*\right)$ curve again uses the corresponding local growth factor concentrations $P\left(x,8\right)$ and $T\left(x,8\right)$. In addition to these curves, we plot individual data points that show the simulated $\left(m,\rho\right)$ values at each $x$ at $t=8$ (open circles), and the maximum $\left(\widehat{m}^*,\widehat{\rho}\,^*\right)$ of each individual $\rho^*\left(m^*\right)$ curve (asterisks). Note, firstly, that the simulated $\left(m,\rho\right)$ data points are all strongly aligned with their corresponding $\rho^*\left(m^*\right)$ curves. This confirms that, at $t=8$, the base case simulation is indeed very close to steady state. Each individual $\rho^*\left(m^*\right)$ curve in Figure~\ref{analytic} has a unique shape and a unique maximum value, but all of the plots share similar qualitative features. In particular, each curve is relatively flat for a wide range of $m^*$ values either side of its maximum at $\widehat{m}^*$. Therefore, when the simulated steady state value of $m$ lies inside the shallow region of the $\rho^*\left(m^*\right)$ curve --- as is the case near the endothelium in the base case simulation --- the simulated steady state value of $\rho$ will differ only slightly from its theoretical maximum. At $x=0$ in the base case simulation, for example, the analytical results give $\widehat{\rho}\,^* \approx 0.415$ and corresponding $\widehat{m}^* \approx 0.155$. The simulated SMC volume fraction at $x=0$ for $t=8$ is much larger ($m\approx0.202)$ than $\widehat{m}^*$, but the corresponding simulated ECM volume fraction ($\rho\approx0.408$) remains very close to $\widehat{\rho}\,^*$. Indeed, Figure~\ref{analytic} indicates that, at $x=0$, any simulated steady state $m$ in the range $[0.102, 0.233]$ would give a steady state $\rho>0.4$ for the same local growth factor concentrations. This suggests that the fibrous cap generated in the base case simulation would be robust to a relatively large reduction in cap region SMC numbers.

\subsubsection{Sensitivity Analysis} \label{sssSens}
The base case simulation results demonstrate how SMCs can accumulate in the plaque and synthesise a collagenous cap in response to diffusible signals from endothelium-derived PDGF and TGF-$\beta$. In this section, we perform additional sensitivity analyses to investigate how the values of certain parameters determine the extent of collagen deposition in the cap region. In particular, we investigate both the impact of SMC adhesion to the nascent collagenous matrix and the role of the relative rates of growth factor influx into the plaque.  
\newline

\noindent \emph{SMC Affinity for Collagen} \\
\emph{In vitro} studies have demonstrated that vascular SMCs can elicit a strong haptotactic response to types I, IV and VIII collagen \citep{Nels96, Hou00}. The extent of any haptotactic contribution to SMC migration during \emph{in vivo} cap formation is unclear, but, since plaque ECM is known to contain both collagen I and collagen VIII \citep{Adig09}, it seems plausible that SMC haptotaxis may play an important role. We investigate this possibility below by examining the sensitivity of the model to the parameter $\chi_\rho$, which quantifies the affinity of the plaque SMCs for the collagenous ECM phase.

\begin{figure}
		\centering		
  		\includegraphics[height=5.8cm]{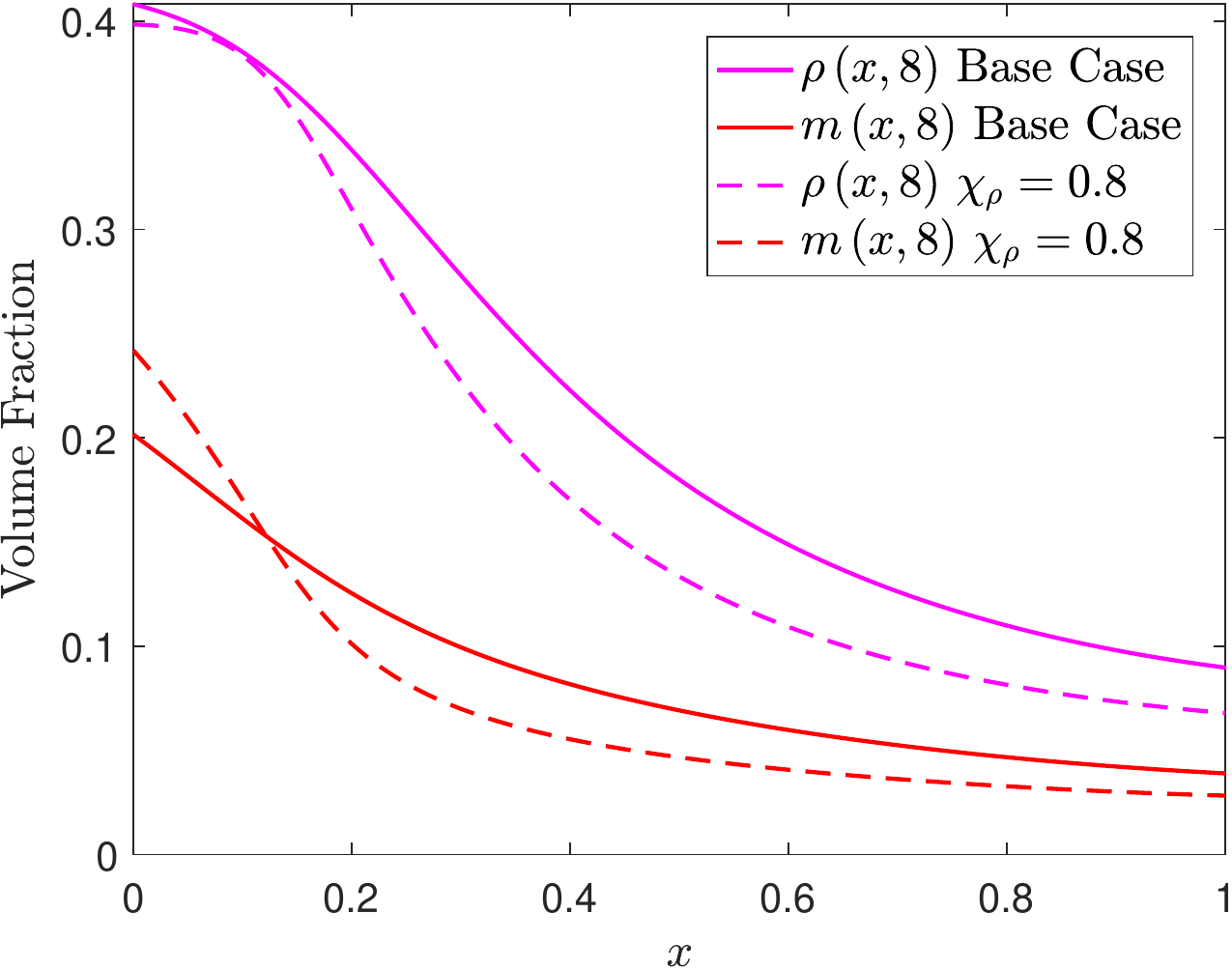}
		\caption{Approximate steady state SMC volume fraction (red lines) and ECM volume fraction (magenta lines) profiles in the plaque for the base case simulation (solid lines) and a simulation with increased SMC affinity for the ECM phase ($\chi_\rho=0.8$; dashed lines). All plots taken at time $t=8$.}\label{mrChirho}
	\end{figure}

We find that setting $\chi_\rho=0$ produces results that are only marginally different to those from the base case simulation. This highlights that our choice of $\chi_\rho=0.3$ for the base case value was particularly conservative, and we therefore consider the impact of a significant increase in this value. For a sensitivity simulation with $\chi_\rho=0.8$, we find that the solution dynamics are practically identical to the base case simulation for the first 5--6 weeks of simulation time. Beyond this point, however, the emerging ECM phase exerts and increasing influence on the SMC behaviour and the model solution for larger $\chi_\rho$ diverges from the base case simulation dynamics. Figure~\ref{mrChirho} compares the approximate steady state SMC and ECM profiles from the base case simulation and from the case with $\chi_\rho=0.8$. As would be expected, the increase in $\chi_\rho$ results in increased recruitment of SMCs to the region proximal to the endothelium. More surprising, however, is that the overall plaque SMC content is reduced from 8.6\% to 7.1\%, and that there is no overall increase in ECM deposition in the cap region. The reduction in SMC numbers appears to be due to an increased squeeze on the PDGF influx from the endothelium, which reduces both SMC recruitment from the media and SMC mitosis inside the intima. The lack of increase in cap ECM deposition, on the other hand, can be attributed to the fact that the SMC volume fraction near the endothelium has increased well beyond the level required to maximise ECM synthesis (c.f.\ Figure~\ref{heat(P,T)}). Of course, a potential benefit of the increased haptotaxis simulated here is that less ECM is deposited \emph{beyond the cap region} (due to the drop in SMC numbers), which may contribute to a reduction in the overall hardening of the diseased arterial tissue.
\newline 

\noindent \emph{Growth Factor Influx} \\
A key target of this study has been to understand how plaque SMC behaviour and fibrous cap formation are regulated by the growth factors PDGF and TGF-$\beta$. Since it is difficult to quantify the concentrations of these chemicals in a given \emph{in vivo} plaque, we perform a range of sensitivity analyses below to investigate how the relative rates of PDGF and TGF-$\beta$ influx from the endothelium can contribute to the density, and therefore the stability, of the collagenous cap. We first examine the consequences of varying the growth factor influx rates $\alpha_T$ and $\alpha_P$ independently, before considering the impact of varying them simultaneously.

To facilitate the comparison of cap formation dynamics in the simulations that we perform in this section, we introduce a simple metric to measure the SMC and ECM levels proximal to the endothelium at a given point in time. Specifically, we define the quantity $V_i\left(t;X\right)$, which denotes the average volume fraction of phase $i$ ($i=m,\rho$) in the interval $x \in \left[0,X\right]$ at time $t$:	
	\begin{equation}
		V_i\left(t;X\right)=\frac{1}{X}\int_{0}^{X} i\left(x,t\right) \mathrm{d} x,
	\end{equation}
where $0<X \ll 1$. In all that follows, we choose (arbitrarily) to set $X=0.2$ and assume that the spatial interval $x \in \left[0,0.2\right]$ represents what we shall refer to as the cap region. We remark that we have calculated $V_i$ values for several different choices of the cap region width, but (within reasonable bounds) the precise choice of $X$ does not alter our qualitative findings in any of the below sensitivity studies.

\begin{figure}
		\centering		
  		\includegraphics[height=5.8cm]{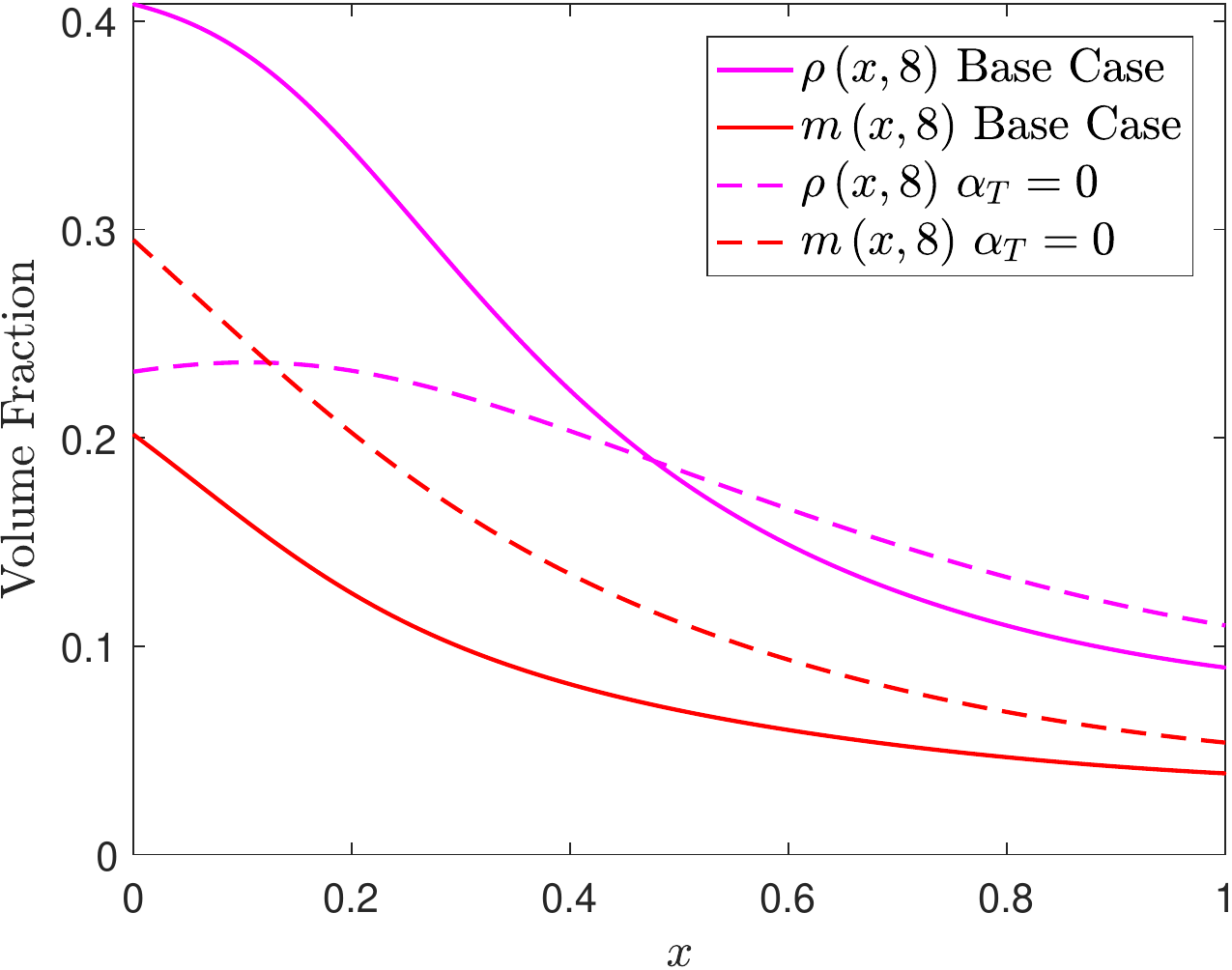}
		\caption{Approximate steady state SMC volume fraction (red lines) and ECM volume fraction (magenta lines) profiles in the plaque for the base case simulation (solid lines) and a simulation with no influx of TGF-$\beta$ ($\alpha_T=0$; dashed lines). All plots taken at time $t=8$.}\label{mrAlphaT}
	\end{figure}

Figure~\ref{mrAlphaT} presents the approximate steady state SMC and ECM profiles for a simulation with no influx of TGF-$\beta$ (i.e.\ $\alpha_T=0$). This scenario mimics the experimental set-ups used by \citet{Mall01} and \citet{Lutg02}, both of whom studied the impact of blocking TGF-$\beta$ signalling during plaque growth in the ApoE knockout mouse. The results show that the absence of TGF-$\beta$ has reduced the total collagen deposition in the plaque by approximately 15\%. This is a smaller reduction in collagen deposition than may have been expected, but it is clear that collagen levels have been rescued to some degree by a corresponding increase in SMC numbers. Indeed, this increase in SMC numbers is directly attributable to the reduced rate of collagen deposition because contact inhibition of SMC proliferation remains lower, and PDGF influx from the endothelium remains higher, throughout the simulation. Having said that, the simulation results still demonstrate that the absence of TGF-$\beta$ can decimate the fibrous cap. Figure~\ref{mrAlphaT} shows that, near steady state, the average ECM volume fraction in the cap region $V_{\rho}\left(t;0.2\right)$ has decreased by approximately 40\% from 0.381 to 0.235. This result is consistent with the findings of \citet{Mall01} and \citet{Lutg02}, who report that inhibition of TGF-$\beta$ signalling can lead to an unstable plaque phenotype.

\begin{figure}
		\centering		
  		\includegraphics[height=5.8cm]{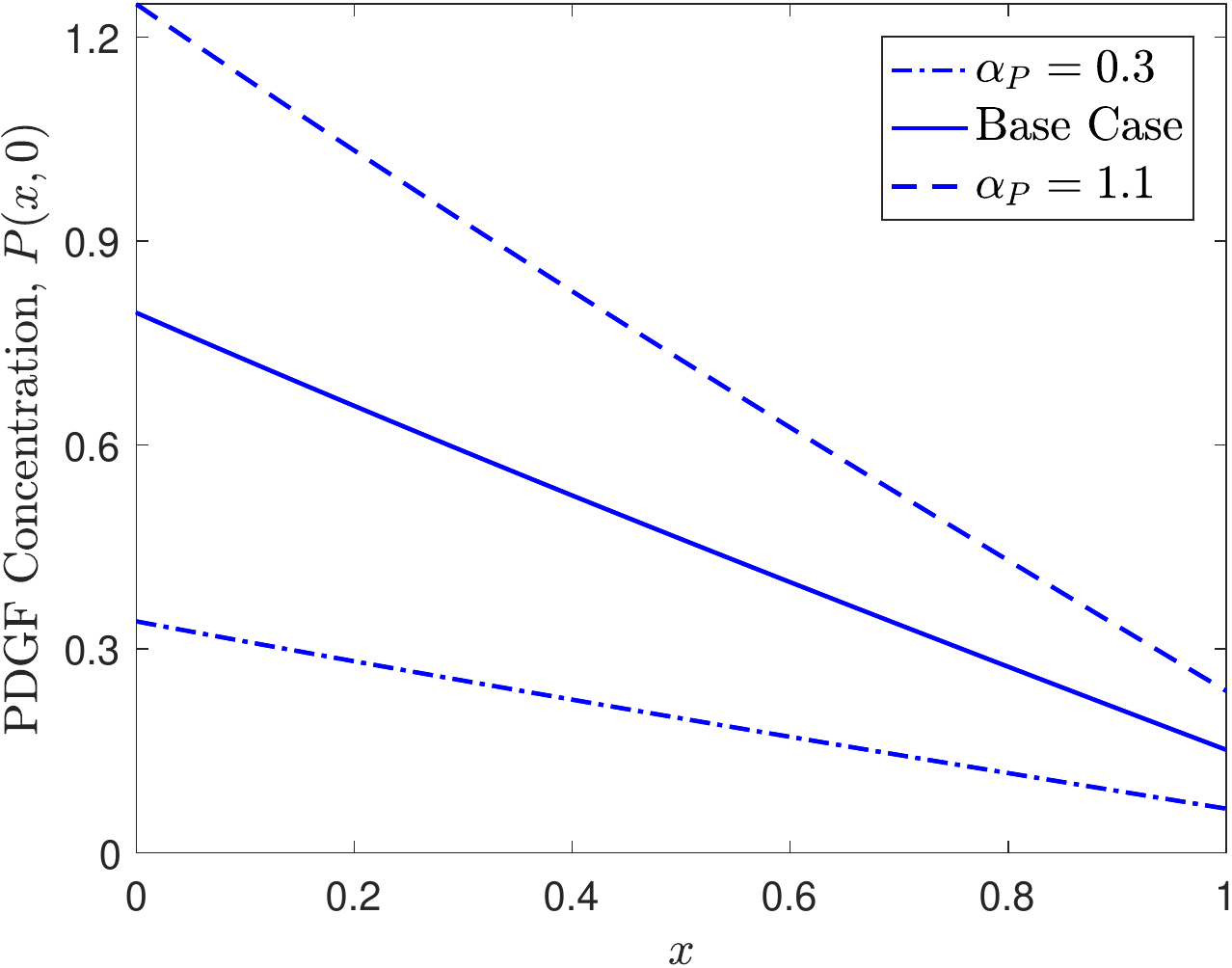}
		\caption{Initial PDGF concentration profiles in the plaque for the base case simulation (solid line), a simulation with a smaller rate of PDGF influx from the endothelium ($\alpha_P=0.3$; dot-dash line) and a simulation with a larger rate of PDGF influx from the endothelium ($\alpha_P=1.1$; dashed line).}\label{initP}
	\end{figure}

\begin{figure}
		\centering
		\begin{subfigure}[b]{0.45\textwidth}
			\includegraphics[width=\textwidth]{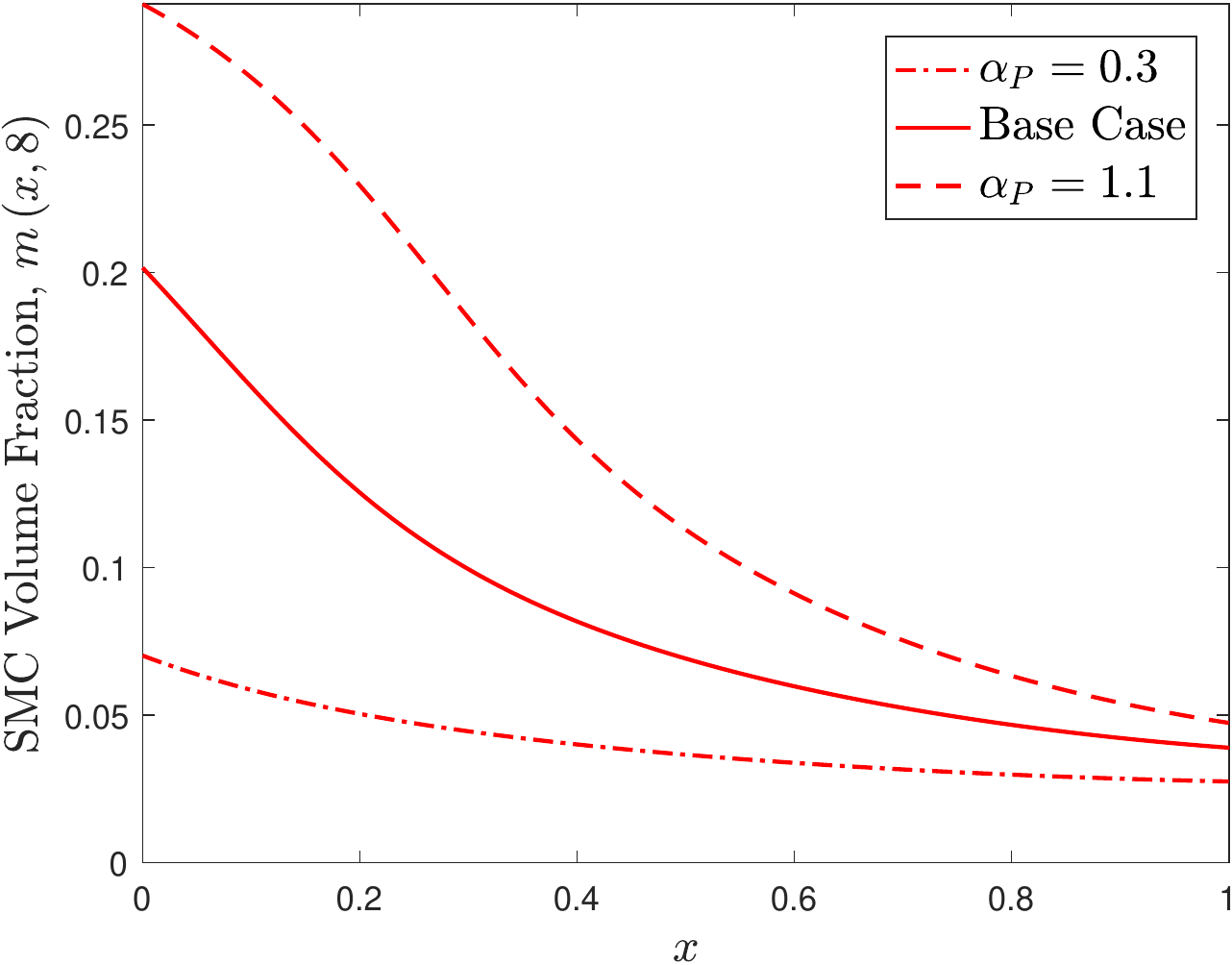}
			\caption{} \label{mAlphaP}
		\end{subfigure}
		\hspace{0.05\textwidth}
		\begin{subfigure}[b]{0.45\textwidth}
			\includegraphics[width=\textwidth]{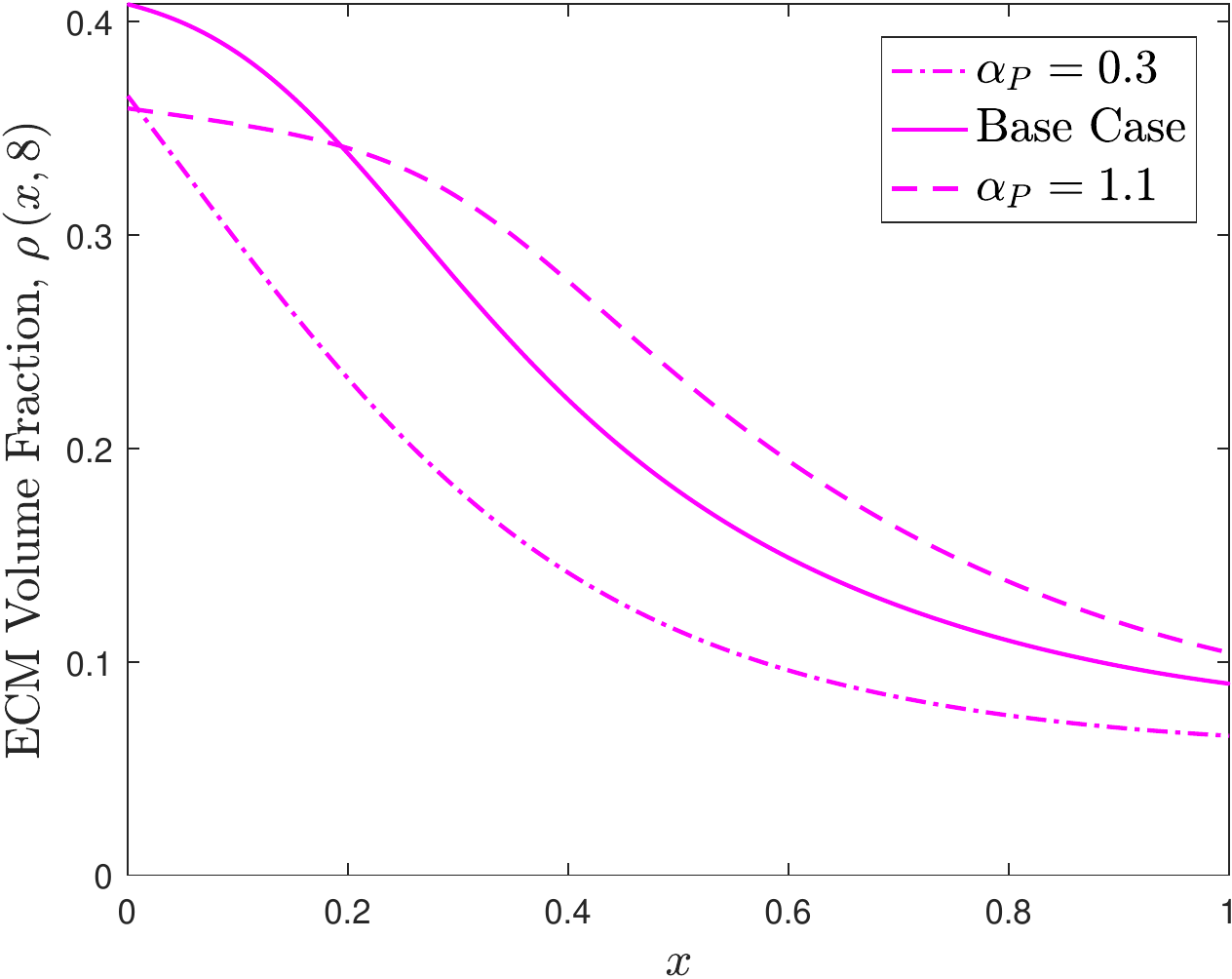}
			\caption{} \label{rAlphaP}
		\end{subfigure}
		\caption{Approximate steady state (a) SMC volume fraction and (b) ECM volume fraction profiles in the plaque for the base case simulation (solid lines), a simulation with a smaller rate of PDGF influx from the endothelium ($\alpha_P=0.3$; dot-dash lines) and a simulation with a larger rate of PDGF influx from the endothelium ($\alpha_P=1.1$; dashed lines). All plots taken at time $t=8$.} \label{mrAlphaP}
	\end{figure}

For PDGF, we cannot simulate a case where chemical signalling is blocked entirely because cap formation in the model is reliant upon PDGF-stimulated recruitment of SMCs from the media. The model does, however, elicit interesting dynamics over a range of non-zero values of $\alpha_P$, and we therefore consider the impact of both an increase and a decrease in this parameter. Figure~\ref{initP} compares the initial PDGF profile from the base case simulation with the initial PDGF profiles for $\alpha_P=0.3$ and $\alpha_P=1.1$. Note that increasing the value of $\alpha_P$ not only increases the PDGF gradient but also increases the PDGF concentration throughout the plaque tissue. Figure~\ref{mrAlphaP} presents the corresponding SMC and ECM  profiles for each value of $\alpha_P$ after 8 months of simulation time. The SMC profiles show a large disparity in their volume fractions, which is particularly pronounced near the endothelium (Figure~\ref{mAlphaP}). This reflects the significant differences in the relative rates of PDGF-stimulated chemotaxis and proliferation in each case. The ECM profiles, on the other hand, show relatively little variation in their volume fractions, but do still demonstrate some interesting features (Figure~\ref{rAlphaP}). For example, the case with larger $\alpha_P$ produces a slightly lower cap collagen content compared to the base case simulation despite recruiting over 60\% more SMCs to the cap region. This result can mostly be attributed to the fact that the cap SMC volume fraction has significantly exceeded the level required to attain optimum ECM synthesis. However, note that the elevated plaque PDGF levels also contribute to an enhanced rate of ECM degradation by increasing MMP production by the invading SMCs. In contrast, in the simulation with smaller $\alpha_P$, the sparse SMC population makes a remarkably strong attempt at cap synthesis. In this case, the collagen content in the cap region is only 20\% less than that in the base case simulation despite a cap SMC content that has been depleted by over 60\%.

\begin{figure}
		\centering
		\begin{subfigure}[b]{0.45\textwidth}
			\includegraphics[width=\textwidth]{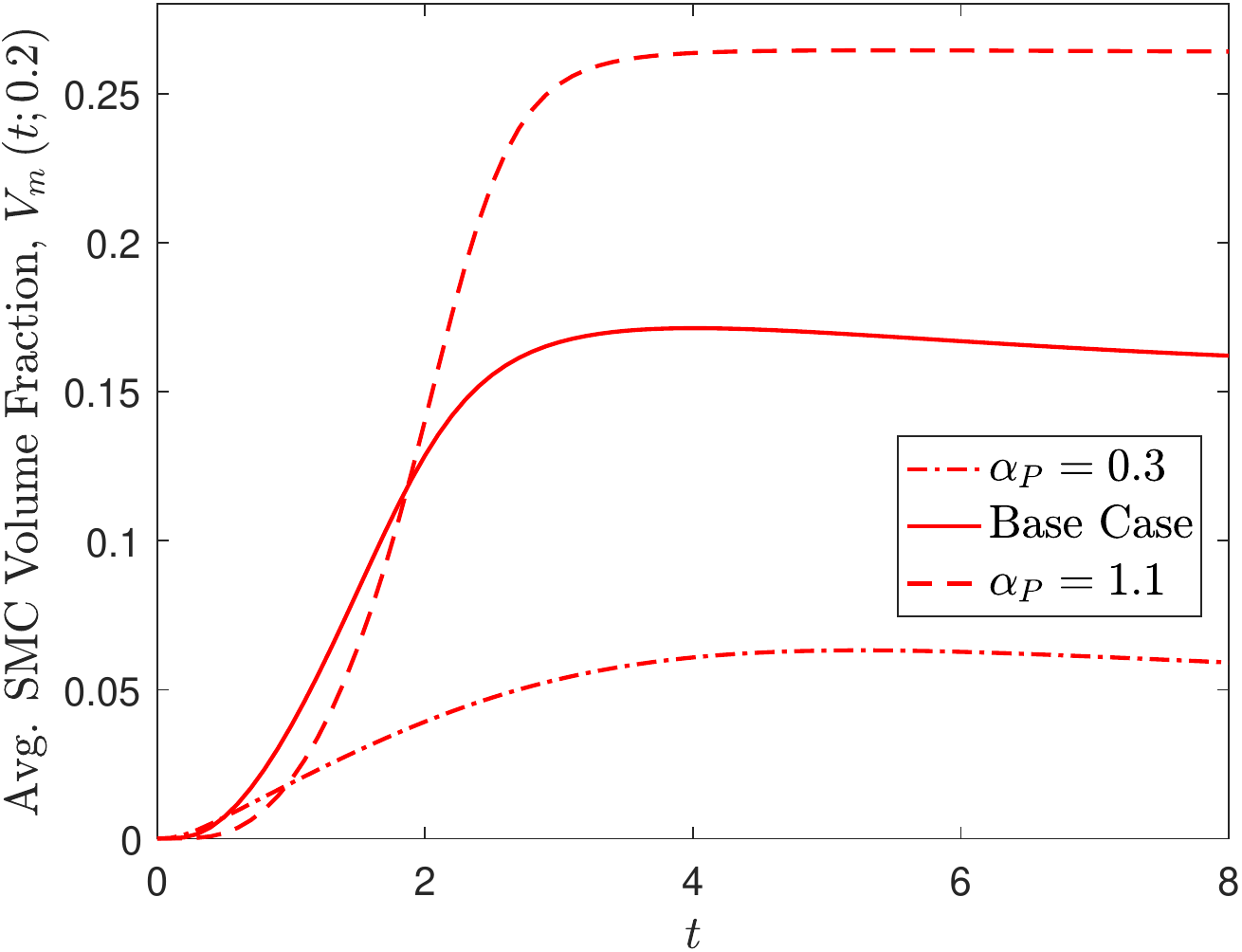}
			\caption{} \label{VmAlphaP}
		\end{subfigure}
		\hspace{0.05\textwidth}
		\begin{subfigure}[b]{0.45\textwidth}
			\includegraphics[width=\textwidth]{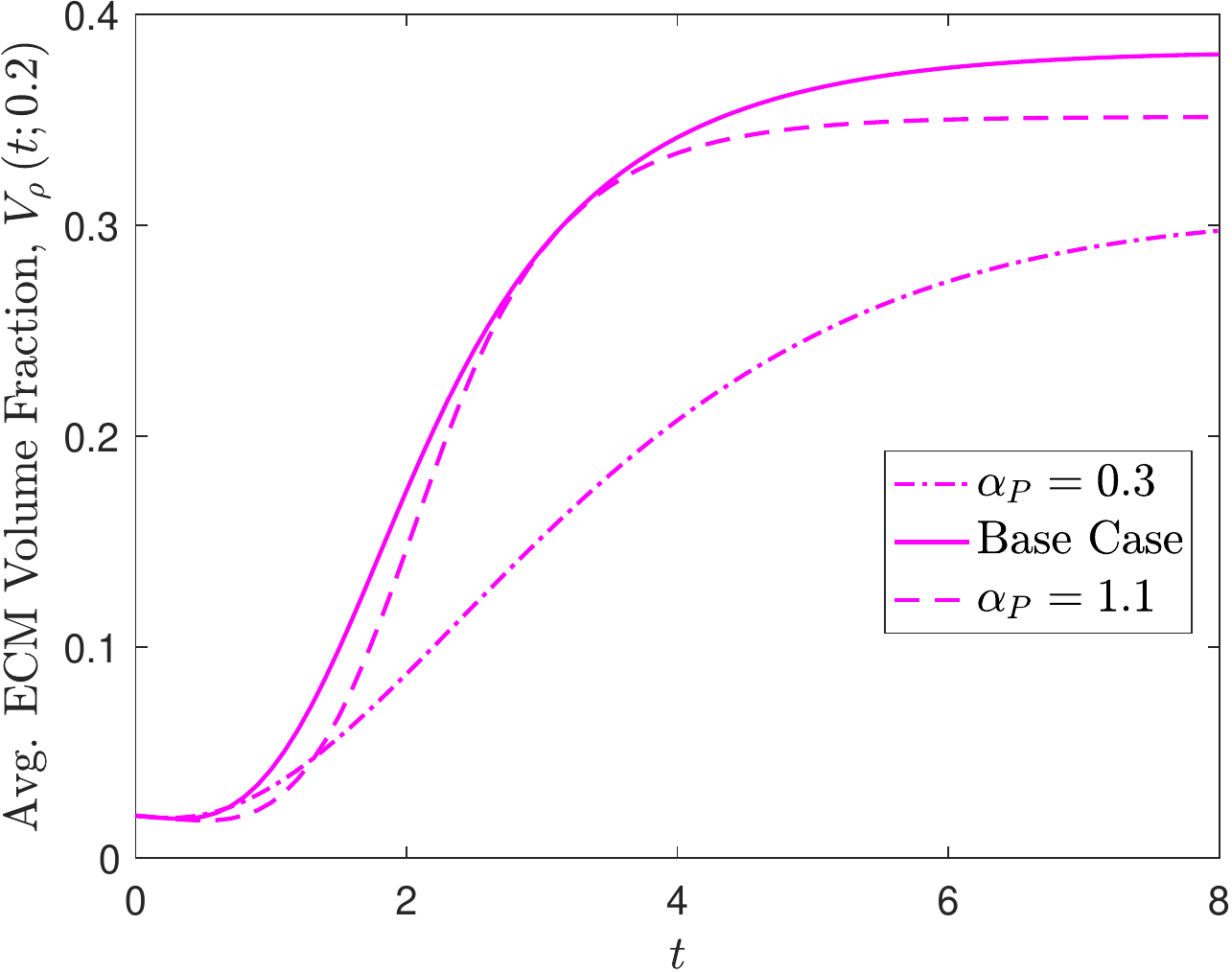}
			\caption{} \label{VrAlphaP}
		\end{subfigure}
		\caption{Plots that show how the average cap region volume fractions of (a) SMCs $V_m\left(t;0.2\right)$ and (b) ECM $V_{\rho}\left(t;0.2\right)$ vary with time for the base case simulation (solid lines), a simulation with a smaller rate of PDGF influx from the endothelium ($\alpha_P=0.3$; dot-dash lines) and a simulation with a larger rate of PDGF influx from the endothelium ($\alpha_P=1.1$; dashed lines).} \label{VmrAlphaP}
	\end{figure}

Of course, if we examine these SMC and ECM profiles at only one time point, we get no real indication of the overall simulation dynamics. Hence, in Figure~\ref{VmrAlphaP}, we also present the average cap SMC volume fraction $V_m\left(t;0.2\right)$ and the average cap ECM volume fraction $V_{\rho}\left(t;0.2\right)$ as functions of time $t$. For $\alpha_P = 1.1$, Figure~\ref{VmAlphaP} shows a slow initial rate of cap SMC recruitment, followed by a period of rapid population growth that saturates after 3--4 months. However, despite a cap SMC population that significantly exceeds the base case levels from 2 months onwards, the temporal accumulation of cap ECM shows a very similar pattern to the base case simulation and fails to exceed the base case value of $V_{\rho}$ at any time (Figure~\ref{VrAlphaP}). For $\alpha_P = 0.3$, however, we observe completely different cap formation dynamics. The reduced PDGF levels in this case inhibit both the rate and the extent of cap SMC recruitment (Figure~\ref{VmAlphaP}), which, in turn, creates a significant lag in the rate of cap ECM accumulation (Figure~\ref{VrAlphaP}). Hence, although these SMCs clearly have the \emph{capacity} to generate large quantities of ECM, their limited numbers ensure that this process takes an extended period of time. 
	
\begin{figure}
		\centering
		\begin{subfigure}[b]{0.48\textwidth}
			\includegraphics[width=\textwidth]{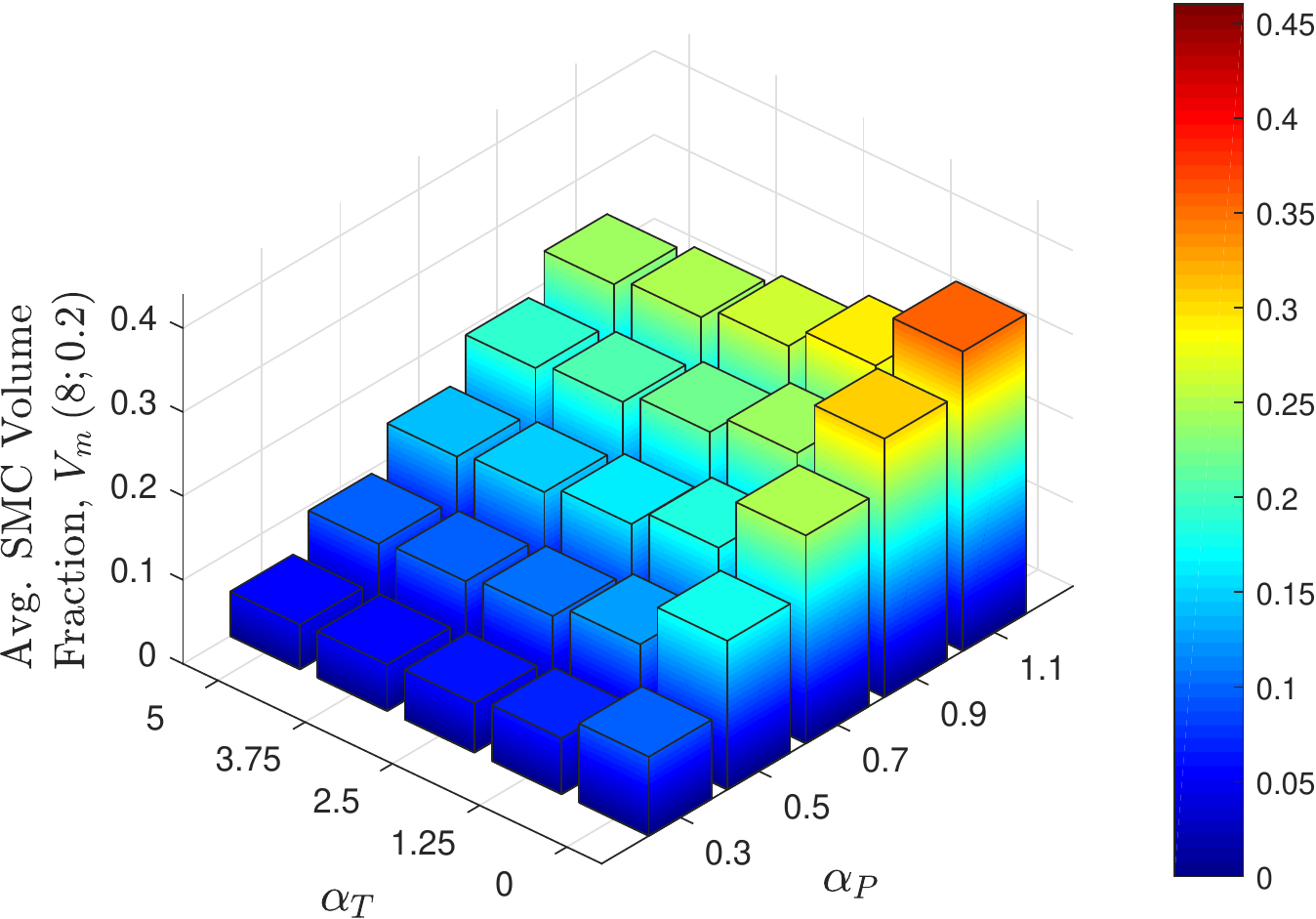}
			\caption{} \label{mGF}
		\end{subfigure}
		\hspace{0.02\textwidth}
		\begin{subfigure}[b]{0.48\textwidth}
			\includegraphics[width=\textwidth]{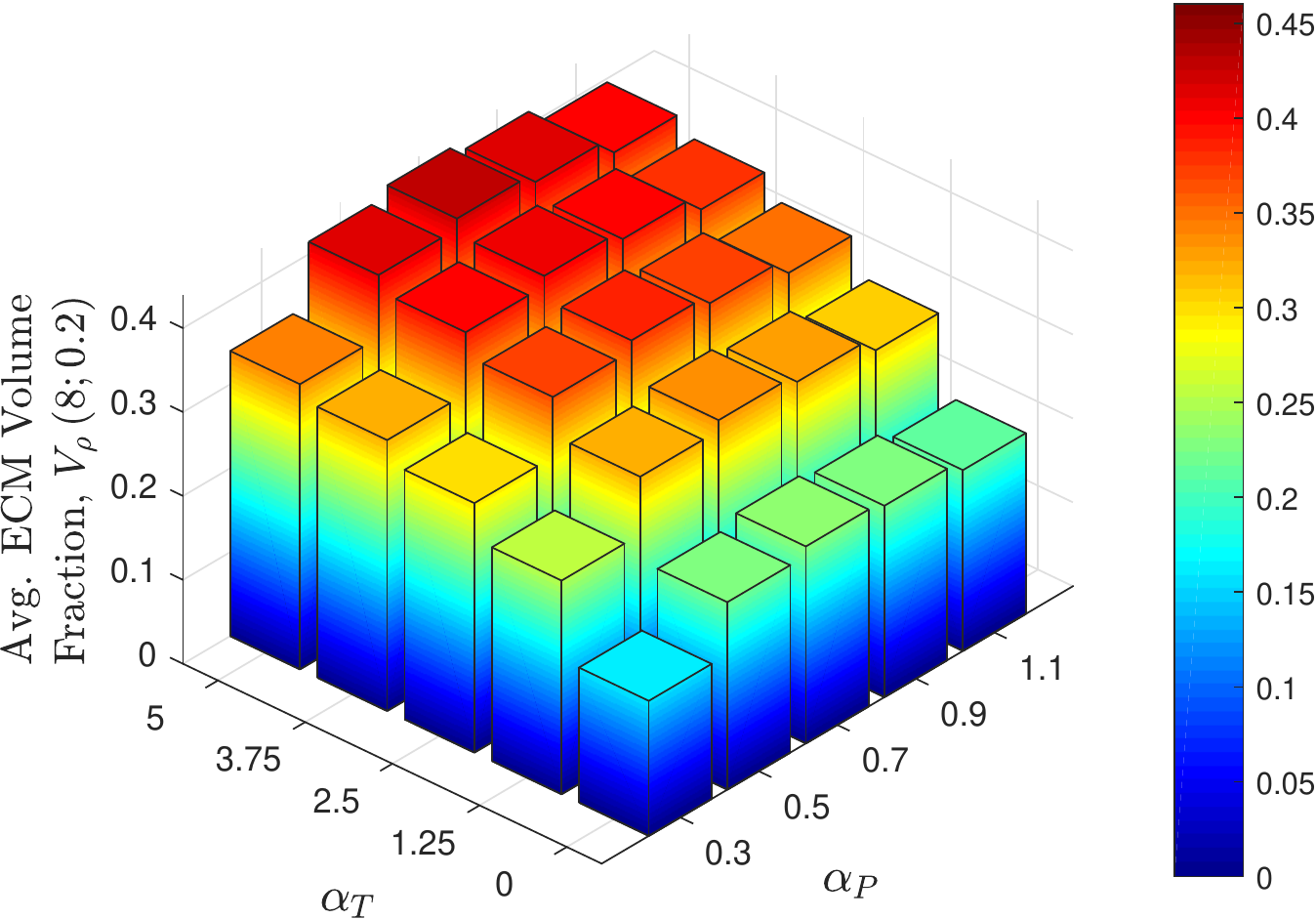}
			\caption{} \label{rGF}
		\end{subfigure}
		\caption{Charts that show how the rates of PDGF influx $\alpha_P$ and TGF-$\beta$ influx $\alpha_T$ influence the average volume fractions of (a) SMCs $V_m\left(t;0.2\right)$ and (b) ECM $V_{\rho}\left(t;0.2\right)$ in the cap region after 8 months of simulation time. Individual simulations were performed for each unique pair of values from the lists $\alpha_P =$ 0.3, 0.5, 0.7, 0.9, 1.1 and $\alpha_T =$ 0, 1.25, 2.5, 3.75, 5.  The central bar on each chart represents the outcome of the base case simulation.} \label{mrGF}
	\end{figure}

We conclude this study by briefly considering the consequences of simultaneously changing the rates of influx of both PDGF and TGF-$\beta$. Figure~\ref{mrGF} presents bar charts of $V_{\rho}\left(t;0.2\right)$ and $V_m\left(t;0.2\right)$ at time $t=8$ for simulations with a range of values of $\alpha_T$ and $\alpha_P$. Specifically, we report results for each unique pair of values from $\alpha_T =$ 0, 1.25, 2.5, 3.75, 5 and $\alpha_P =$ 0.3, 0.5, 0.7, 0.9, 1.1. Note that results from the base case simulation are represented by the central bar on each chart. Figure~\ref{mGF}, which plots the $V_m$ value for each simulation, shows two clear trends: (1) increasing $V_m$ with increasing $\alpha_P$ (mainly due to increased SMC proliferation) and (2) decreasing $V_m$ with increasing $\alpha_T$ (due to the inhibitory effects of increased ECM deposition). Interestingly, the plot for $V_{\rho}$ indicates a relatively large region of the $\left(\alpha_T,\alpha_P\right)$ parameter space where the model predicts a robust cap formation response (Figure~\ref{rGF}). This region corresponds to parameter regimes with large TGF-$\beta$ levels, moderate PDGF levels and, consequently, moderate volume fractions of cap SMCs.
	
\section{Discussion} \label{sDiscuss}
Mature atherosclerotic plaques contain a lipid-rich core of necrotic material and release of this material into the circulation can cause myocardial infarction or stroke. Fibrous cap formation provides protection against these outcomes by stabilising the plaque and sequestering the dangerous plaque content from the bloodstream. However, the mechanisms that regulate cap formation, and the factors that may subsequently lead the cap to fail, are poorly understood. In this paper, we have developed a multiphase model to investigate collagenous cap synthesis by growth factor-stimulated SMCs. The model includes representations of the growth factors PDGF and TGF-$\beta$, and we have used the model to study the co-operation (and competition) between these diffusible chemicals in the recruitment of vascular SMCs to remodel the plaque ECM.

\subsection{Applicability of the model to human atherosclerosis} \label{}
The model that we have developed in this study is designed to investigate cap formation in the ApoE-deficient mouse. We have chosen to focus on this transgenic mouse model of atherosclerosis because there exists an extensive experimental literature that we have used to inform and validate our modelling assumptions. It would, of course, be possible to extrapolate our model to human atherosclerosis by solving the equations on a domain of significantly greater dimensional width (e.g. 1~\si{\milli\metre}). However, the behaviour of vascular SMCs in human plaques remains poorly understood \citep{Benn16}. Unlike mice, healthy human arteries contain a resident population of intimal SMCs and the extent of medial SMC recruitment to human plaques has yet to be established.

\subsection{Comparison with previous work and limitations of the model} \label{}
The research presented in this paper has some similarities and several key differences to the earlier approach of \citet{Wats18}. In \citet{Wats18}, we established a novel modelling framework with non-standard boundary conditions to allow us to study the chemotactic recruitment of plaque SMCs from the media in response to an influx of endothelium-derived PDGF. We did not explicitly model the deposition of a fibrous cap. Instead, we assumed that the plaque ECM profile would evolve to more-or-less reflect the plaque SMC profile and, hence, that an increase in SMC recruitment to the cap region would generally improve cap stability. The current work uses the modelling framework of \citet{Wats18} as a foundation, but builds a substantial new model that focusses on ECM synthesis and degradation by plaque cells in response to an influx of both PDGF and TGF-$\beta$ from the endothelium. The new model also includes haptotactic SMC migration in response to dynamically varying gradients in the ECM and, hence, to the best of our knowledge, this paper presents the first attempt to include both chemotactic and haptotactic cell movement in a multiphase model. The results that we have reported have demonstrated that the plaque SMC profile does not necessarily provide a reliable indication of the corresponding ECM profile, nor of the overall likelihood of fibrous cap stability. This is in contrast to the assumptions that we made previously in the simpler model of \citet{Wats18}.

By retaining the modelling framework from our previous study, we have also retained some of the limitations of the earlier approach. In particular, the model remains on a fixed domain and does not explicitly account for the lipoprotein influx and the associated inflammatory response that may contribute to continued plaque growth during the process of cap formation. The current work can therefore be interpreted as a model of cap formation in a plaque that otherwise exists in a dynamic equilibrium (i.e.\ where any influx of LDL and immune cells from the bloodstream is exactly balanced by an efflux of lipid-loaded foam cells to the adventitial lymphatics). Of course, even with this interpretation of the model, it is possible to argue that we should incorporate domain growth to account for the accumulation of SMCs and collagenous ECM in the plaque. In the numerical results presented in Section~\ref{ssNumSim}, for example, the total fraction of the plaque occupied by SMCs and ECM increases over the course of a simulation by between 17--35\% depending on the choice of parameters. In the absence of domain growth, we assume that this increase in the plaque SMC and ECM levels is balanced by an equivalent reduction in the volume fraction of the generic tissue phase. This reduction in generic tissue phase material can be interpreted as a loss of the initial non-collagenous ECM, which is degraded and subsequently recycled, and as a loss of interstitial fluid, which is squeezed out of the plaque or absorbed by SMCs to produce new material. We note, however, that this interpretation becomes less reasonable if the total SMC and ECM volume fractions in the plaque become large because significant quantities of other generic tissue phase constituents (e.g.\ immune cells, lipids) are also effectively lost from the tissue during cap formation. Despite these limitations of the approach, we believe that the fixed domain model provides a useful and computationally straightforward approximation to the problem of modelling atherosclerotic cap growth. In future studies, we will develop a comprehensive moving boundary approach to explore how dynamic expansion of the intima by immune cell and SMC activity can influence the efficacy of cap formation.

\subsection{Analytical results explain the (local) SMC-ECM relationship} \label{}
One of the key benefits of employing a fixed domain is that the model remains amenable to mathematical analysis. In Section~\ref{ssAnalyt}, we used the absence of a flux term in the ECM phase equation to study the factors that determine the steady state ECM volume fraction at a given position in the plaque. Specifically, we showed, for fixed concentrations of the growth factors PDGF and TGF-$\beta$, that the steady state ECM volume fraction $\rho^*$ has a biphasic dependence on the steady state SMC volume fraction $m^*$, and, correspondingly, that there exists a steady state SMC volume fraction $\widehat{m}^*$ that gives a maximum steady state ECM volume fraction $\widehat{\rho}\,^*$. Moreover, we derived simple expressions for $\widehat{m}^*$ and $\widehat{\rho}\,^*$ in terms of the dimensionless parameter groupings $\mu$ and $\lambda$, where $\mu$ represents the ratio of the net rate of ECM synthesis by SMCs to the net rate of ECM degradation by immune cells and $\lambda$ represents the ratio of the net rate of ECM degradation by SMCs to the net rate of ECM degradation by immune cells. The most significant finding of this analysis is that, for parameter values informed by a range of \emph{in vitro} and \emph{in vivo} studies, the model predicts that the maximum ECM volume fraction can generally be achieved by a relatively small volume fraction of SMCs (e.g.\ $<20\%$ even at moderate TGF-$\beta$ concentrations). Interestingly, for several of the numerical simulations presented in Section~\ref{ssNumSim}, the model predicts steady state SMC volume fractions proximal to the endothelium that are \emph{above} the level required to maximise ECM deposition. Consequently, the steady state ECM levels in the cap region in these simulations are typically smaller than their maximum possible level. With regard to plaque stability, this outcome initially appears to be undesirable. However, note that a slight excess in SMC numbers above the optimum level may be beneficial because it affords the plaque greater robustness to destabilisation by mechanisms such as SMC loss or diminished TGF-$\beta$ signalling. When the typical steady state SMC volume fraction in the cap region is \emph{below} the level required for maximum ECM deposition, then an equivalent drop in SMC numbers or loss of TGF-$\beta$ signalling would have far more pronounced consequences for plaque stability.

\subsection{Numerical results agree with experimental observations} \label{}
The base case simulation results reproduced several observations from experimental studies of plaque growth in the ApoE-deficient mouse. For example, temporal changes in the model plaque SMC and collagenous ECM levels were consistent with the qualitative pattern reported in \citet{Reif12}. The plaque SMC content accumulated rapidly over the first 2 months of simulation time and maintained a relatively stable level thereafter. The plaque ECM content, on the other hand, accumulated more slowly and eventually reached a stable level after 4--6 months. The total SMC content and the total ECM content in the model plaque after 4--8 months of simulation time were in good quantitative agreement with the values reported in \citet{Reif12} and in other experimental studies \citep{Mall01, Clar06}. An interesting aspect of the base case simulation was that the total plaque SMC content reached a peak of around 9.2\% after 4 months of simulation time and then dropped to around 8.6\% as the plaque ECM content continued to increase. This is a relatively subtle effect, but for simulations where the overall plaque ECM content reaches higher levels (e.g.\ sensitivities with larger $r_s$, smaller $r_d$ or smaller $\beta_\rho$), we observe a more pronounced suppression of SMC numbers in the latter stages of cap formation (results not shown). These findings are consistent with observations made by \citet{Fuku04}, who studied plaque growth in ApoE mice that were resistant to degradation of type I collagen. \citet{Fuku04} found that, relative to standard ApoE mice, these compound-mutant mice had significantly more collagen and significantly fewer SMCs in their plaques. Moreover, the plaques that were resistant to collagen degradation also showed evidence of reduced SMC proliferation and increased SMC death. These observations suggest that excessive accumulation of collagen in plaques can tip the balance towards a net reduction in plaque SMC levels. The model presented here makes a similar prediction and suggests that the loss of SMCs from collagen-rich plaques can be explained by an increase in ECM-mediated contact inhibition of SMC proliferation and a decrease in the capacity for PDGF transport in the dense intimal tissue.

A consistent finding throughout this modelling study is that the growth factor TGF-$\beta$ plays a critical role in the ability of plaque SMCs to deposit a stable, collagen-rich fibrous cap. The analytical results in Section~\ref{ssAnalyt}, for example, indicated that the presence of TGF-$\beta$ can enable significantly fewer SMCs to deposit significantly more ECM than would otherwise be possible. These analytical results have been further supported by numerical simulations. Compared to the base case, a sensitivity simulation with no TGF-$\beta$ ($\alpha_T=0$) showed an overall decline in steady state plaque ECM levels, which included a substantial reduction in ECM deposition in the cap region. These observations are qualitatively consistent with those from equivalent experimental studies in ApoE mice \citep{Mall01, Lutg02}. However, we also note some inconsistencies between the \emph{in vivo} and \emph{in silico} results. For example, the model predicted a 15\% decrease in plaque ECM content coupled to a 55\% increase in plaque SMC content, whereas both of the above experimental studies showed a 50\% decrease in plaque collagen and no overall change in plaque SMC content after TGF-$\beta$ blockade. The reasons for these differences are not clear, but the experimental results also showed that inhibition of TGF-$\beta$ signalling led to increased inflammatory cell content and larger lipid cores in the murine plaques. The SMC response in these \emph{in vivo} plaques may therefore have been blunted by increased competition for space or by increased SMC death in the noxious and highly inflammatory plaque environment. The absence of these confounding factors in the model could potentially explain why the plaque SMC and ECM content are over-predicted in the current \emph{in silico} approach.

In order for TGF-$\beta$ to play its role in cap formation, PDGF is first required to stimulate SMC recruitment to the cap region. PDGF may contribute both positively and negatively to cap formation. Experimental evidence suggests that PDGF can increase SMC production of matrix-degrading MMPs, in addition to stimulating SMC migration and SMC mitosis. These factors were included in the model, and sensitivity simulations with different rates of PDGF influx from the endothelium $\alpha_P$ showed interesting results. For $\alpha_P$ values in the range 0.3--1.1, simulations showed that SMC recruitment to the cap region increased significantly with increasing $\alpha_P$. ECM levels in the cap region close to steady state, on the other hand, were relatively insensitive to $\alpha_P$ and showed a biphasic response, with maximal ECM deposition at the base case value ($\alpha_P=0.7$). Note that these qualitative patterns of SMC and ECM sensitivity to changes in $\alpha_P$ have been shown to be independent of the corresponding rate of TGF-$\beta$ influx (for $\alpha_T$ in the range 0--5).

The reported simulation with $\alpha_P=0.3$ (and $\alpha_T$ at its base case value) is interesting because it demonstrates that even a small amount of SMCs can generate a substantial amount of ECM. From a total plaque SMC volume fraction of only 4.0\% (vs.\ 8.6\% in the base case simulation), the total plaque ECM volume fraction reaches 15.0\% (vs.\ 21.4\% in the base case simulation). This result is qualitatively consistent with observations made by \citet{Clar06} in a study that examined the impact of diphtheria toxin (DT)-induced SMC apoptosis in the plaques of ApoE mice. Plaque SMC content in DT treated mice was reduced more than 4-fold relative to untreated mice (2.5\% vs.\ 10.2\%), but the plaque collagen content was correspondingly reduced by just over 2-fold (8.0\% vs.\ 18.1\%). Of course, despite the relatively small drop in the long-term plaque collagen content, the simulation with $\alpha_P=0.3$ also showed a significant delay in cap formation relative to the base case. This simulation therefore provides a plausible explanation for the observations of \citet{Koza02}, who studied the influence of PDGF blockade on cap formation in ApoE mice. In this experimental study, two different techniques were used to inhibit PDGF signalling: (1) development of transgenic ApoE mice with no PDGF expression in their circulating cells and (2) treatment of standard ApoE mice with a PDGF receptor antagonist. The authors indicated that PDGF signalling may only have been partially disrupted in each case, and both methods were shown to delay rather than prevent cap formation. \citet{Koza02} did not report any data on the SMC content of the plaques in their study, but the reported influence of PDGF inhibition on \emph{in vivo} fibrous cap formation is entirely consistent with the modelling results presented in this paper.

\subsection{SMC haptotaxis and the problem of model ill-posedness} \label{}
SMCs are well-known to respond haptotactically to gradients in collagenous substrates and we have included this phenomenon in the model by assuming that the SMC phase has an affinity for the ECM phase (quantified by the parameter $\chi_\rho$). Unlike SMC chemotaxis, haptotactic SMC migration is \emph{not} critical to the formation and maintenance of a cap in the model. We therefore assumed a relatively low base case value for $\chi_\rho$, and later performed a sensitivity study to investigate the impact of an increase in this value. The sensitivity study emphasised the importance of SMC chemotaxis on the initiation of cap formation because, even with a relatively large $\chi_\rho$ value, it took over a month of simulation time for the nascent ECM profile to have a discernible influence on the SMC phase. In the long term, the increase in SMC haptotaxis led to steeper profiles in both the SMC and ECM phases in the plaque but no overall increase in ECM deposition in the cap region. It would be interesting to perform a more comprehensive investigation of the impact of the relative contributions of SMC haptotaxis and SMC chemtotaxis on fibrous cap formation by simultaneously varying the values of $\chi_\rho$ and $\chi_P$ (or $\alpha_P$). An in-depth investigation of the role of SMC haptotaxis was not the primary intention of the current work, but the model presented herein creates the opportunity for a more focussed future study.

One reason why a thorough study of the role of SMC haptotaxis is challenging is that the model equations can become ill-posed if the value assigned to $\chi_\rho$ is too large. To understand why this is the case, consider the form of the effective SMC diffusion coefficient in equation~(\ref{mmasnd}). This can be shown to be proportional to:
	\begin{equation}
		\Lambda + \rho \left(\psi + m \frac{\partial \psi}{\partial m}\right) \, = \, \frac{\chi_P}{1 + {\left(\kappa P \right)}^{n_P}} \, - \, \rho \chi_\rho \, + \, \frac{\delta \rho m^{n_\rho} \big[ \left(1+n_\rho\right)\left(1-\rho\right)-m \big]}{\left(1-m-\rho\right)^{n_\rho+1}}. \label{effDiff}
	\end{equation}
Note that the third term on the right hand side of this expression is strictly positive because $n_\rho>0$ and, by the no voids condition of equation~(\ref{nvoi}), $m<1-\rho$. Recall that we have assumed zero flux for SMCs at the endothelium (equation~(\ref{mBC0})). This boundary condition stipulates that, at $x=0$, any SMC flux towards the endothelium must be exactly balanced by an equivalent SMC flux in the opposite direction. In practice, the chemotactic and haptotactic SMC fluxes almost always act towards the endothelium and this creates a negative gradient in the SMC phase at $x=0$. Hence, to satisfy the zero flux boundary condition at $x=0$, the diffusive SMC flux must always be positive and act down the SMC gradient away from the endothelium. If the diffusive SMC flux becomes negative at any stage, the problem immediately becomes ill-posed. Inspection of equation~(\ref{effDiff}) indicates that negative SMC diffusion can occur if the second term on the right hand side is dominant. This can occur if SMC adhesion to the collagenous ECM is sufficiently strong to counteract the other mechanisms of cell spreading. Note that this scenario is particularly likely to occur if $P$ is large (first term on the right hand side of equation~(\ref{effDiff}) negligible) and/or if $m$ is small (third term on the right hand side of equation~(\ref{effDiff}) negligible). In general, it is difficult to remove this problem of ill-posedness entirely, but here we propose several possible strategies to at least reduce the region of parameter space where the problem occurs. First, we could regularise the governing equations by including viscous effects in equation~(\ref{taum}) for the SMC phase. Second, as done by \citet{Byrn04}, we could introduce a constant background SMC diffusion by including an additional constant SMC phase pressure in the function $\Lambda$. Third, we could assume that the SMC phase affinity for the ECM phase decreases with increasing PDGF concentration (i.e.\ $\chi_\rho \equiv \chi_\rho\left(P\right)$). In this latter case, the first and second terms on the right hand side of equation~(\ref{effDiff}) would become complementary and, since we also assume that SMC proliferation increases with increasing $P$, this would represent the introduction of a ``go-or-grow'' mechanism in the SMC phase. Note, however, that the inclusion of a $P$ dependency in $\chi_\rho$ would also introduce a new term into the effective SMC chemotaxis coefficient. This term would resemble an ECM phase-dependent chemorepulsion. Of course, the downside of these potential solutions to the problem of ill-posedness is that each solution would introduce additional model parameters whose values may be difficult to determine in practice.

\subsection{Implications for plaque destabilisation mechanisms} \label{}
So, what conclusions can we draw from the model about factors that have been reported to contribute to plaque instability? The model results suggest that, under conditions where initial SMC recruitment is strong and plaque TGF-$\beta$ levels remain favourable, the fibrous cap should be robust to a relatively significant loss of SMC numbers. However, if as suggested by \citet{Wang15} that long-term loss of SMCs can occur due to replicative senescence, the model would predict rapid degradation of the cap by immune cell activity as the SMC population falls to critically low levels. In conditions where the SMC numbers and the TGF-$\beta$ concentration in the plaque remain at moderate levels, the current model suggests that an increase in ECM degradation by immune cells alone is unlikely to destabilise the plaque. In results that we have not reported in this paper, a sensitivity simulation with a 2-fold increase in the baseline rate of ECM degradation by immune cells ($\beta_\rho=1.5$) predicts only a relatively small change in the ECM volume fraction in the cap region (32\% vs.\ 38\% in the base case simulation). This makes sense because, based on our modelling assumptions, the majority of the plaque immune cell content resides deeper in the plaque and beneath the cap region. Of course, the current model is unable to predict the consequences of immune cell ECM degradation at the plaque shoulders, where rupture events are commonly reported to occur \citep{Benn16}. Investigation of this phenomenon will require models for cap formation to be developed in two or three spatial dimensions.

The most critical factor in maintenance of plaque stability, as predicted by the current model, is the continued presence of plaque TGF-$\beta$ and, more importantly, the continued capacity for stimulation of SMCs (and immune cells) by this growth factor. Our analysis and our numerical simulations have indicated that TGF-$\beta$ can play a critical role not only in the formation of a stable cap region, but also in ensuring that cap formation is efficient and does not require excessive recruitment of SMCs. Interestingly, recent reports from \emph{in vivo} and \emph{in vitro} experiments have shown that cholesterol loading in several cell types, including vascular SMCs, can significantly attenuate cellular responsiveness to TGF-$\beta$ \citep{Chen07, Veng15}. Given the extent of cholesterol accumulation in the intimal tissue during atherosclerotic plaque growth, it is therefore plausible that modified LDL consumption by SMCs (and by macrophages) indirectly contributes to the breakdown of plaque ECM by inhibiting the protective actions of TGF-$\beta$. Investigation of this hypothesised cap destabilisation mechanism is another important target for future modelling studies.

\section{Conclusions}
In this paper, we have developed a novel three-phase model of fibrous cap formation in the atherosclerosis-prone mouse. The model investigates the roles of endothelium-derived PDGF and TGF-$\beta$ in the regulation of collagen cap deposition by synthetic vascular SMCs, which migrate both chemotactically and haptotactically in the arterial intima. We have parameterised the model using data from a wide range of \emph{in vitro} and \emph{in vivo} studies and our numerical simulations reproduce a number of experimental observations. Our results provide some interesting insights into potential mechanisms of plaque instability and long-term cap degradation. The model presented in this paper can be extended in several new directions, and this work therefore represents an important step in the development of a dynamic and mechanistic understanding of atherosclerotic plaque formation.

\section*{Acknowledgements}
MGW, CM and MRM acknowledge funding from an Australian Research Council Discovery Grant (DP160104685).

\bibliography{Watson_SMC_New}

\end{document}